\documentclass[twocolumn]{aastex63}

\usepackage{graphicx}
\usepackage{epsfig}
\usepackage{rotating}
\usepackage{natbib}
\usepackage{hyperref}[colorlinks=true, allcolors=blue]

\revised{\today}

\submitjournal{ApJ}

\shorttitle{Comparison of MSPs in Globular Clusters and Galactic Field}
\shortauthors{Lee et al.}

\begin{document}

\title{A comparison of millisecond pulsar populations between globular clusters and the Galactic field}


\author{Jongsu Lee}
\affiliation{Department of Astronomy, Space Science and Geology, Chungnam National University, Daejeon 34134, Republic of Korea}

\author{C. Y. Hui}
\affil{Department of Astronomy and Space Science, Chungnam National University, Daejeon 34134, Republic of Korea}

\author{J. Takata}
\affil{Institute of Particle Physics and Astronomy, Huazhong University of Science and Technology, People`s Republic of China}

\author{A. K. H. Kong}
\affil{Institute of Astronomy, National Tsing Hua University, Hsinchu 30013, Taiwan}

\author{Pak-Hin Thomas Tam}
\affil{School of Physics and Astronomy, Sun Yat-sen University, Guangzhou 510275, People’s Republic Of China}

\author{Kwan-Lok Li}
\affil{Department of Physics, National Cheng Kung University, 70101 Tainan, Taiwan}

\author{K. S. Cheng}
\affil{Department of Physics, University of Hong Kong, Pokfulam Road, Hong Kong}


\begin{abstract}
We have performed a systematic study of the rotational, orbital and X-ray properties of millisecond pulsars (MSPs) in globular clusters (GCs) and compared their nature with those of the MSPs in the Galactic field (GF). We found that GC MSPs generally rotate slower than their counterparts in the GF. Different from the expectation of a simple recycling scenario, no evidence for the correlation between the orbital period and the rotation period can be found from the MSP binaries in GCs. There is also an indication that the surface magnetic field of GC MSPs are stronger than those in the GF. All these suggest dynamical interactions in GCs can alter the evolution of MSPs/their progenitors which can leave an imprint on their X-ray emission properties. While the MSPs in both GF and GCs have similar distributions of X-ray luminosity and hardness, our sample supports the notion that these two populations follow different relation between the X-ray luminosity and spin-down power. We discuss this in terms of both pulsar emission model and the observational bias.
\end{abstract}


\section{Introduction}
There is a consensus that millisecond pulsars (MSPs) are formed through the angular momentum transfer from their binary companions \citep{Alpar1982.300,Radhakrishnan...Srinivasan1982.51,Fabian1983.301}. 
The first MSP, PSR B1937+21 was discovered by \cite{Becker1982.300}. 
In comparison with the non-recycled canonical pulsars, MSPs are characterized by their fast rotation ($P\lesssim20$~ms) and weak surface magnetic fields ($B_{s}\lesssim10^{9}$~G). 
Thanks to the extensive surveys and the synergy of multiwavelength observations \citep[see][for a review]{Hui2018.51} , the currently known MSP population has reached a size of $\sim 600$. 

X-ray emission of MSPs are believed to be originated from the backflow charged particles from the acceleration regions in their magnetospheres \citep[e.g.][]{Zhang...Cheng2003.398}. While the relativistic electron/positron cascades emit the non-thermal synchrotron X-rays when they gyrate along the magnetic field lines, thermal X-ray emission can also be generated when these energetic particles follow the open magnetic field lines and deposit their energies on the stellar surface \citep[e.g.][]{Zavlin2007.308,Bogdanov2008.983}. For the MSPs reside in the compact binaries, additional X-ray emission component can be resulted from the intrabinary shock \citep[e.g.][]{Huang2012.760,Hui2014.781L}.

According to their locations in our Galaxy, MSPs can be divided into two groups: the Galactic field (GF) population and globular cluster (GC) population. For the X-ray properties of MSPs in the GF, \cite{Lee2018.864} have conveyed a systematic survey.  With a left-censored sample of 47 detections and 36 upper limits of their X-ray luminosities $L_{x}$, 
an empirical relation between $L_{x}$ and the spin-down power $\dot{E}$ has found to be $L_{x}\simeq10^{31.05}\left(\dot{E}/10^{35}\right)^{1.31}$~erg~s$^{-1}$ in 2-10 keV. The inferred X-ray conversion efficiency is lower than the previous estimate in the same energy band \citep[e.g.][]{Possenti2002.387} which was subjected to selection bias with the upper-limits excluded in the previous works. 

The X-ray properties of different types of MSPs in the GF have also been compared by \cite{Lee2018.864}. The X-ray emission from the redbacks (RBs), which are characterized by their tight orbits with orbital period $P_{b}\lesssim1$~day and their non-degenerate late-type companions \citep[see][for an updated review]{Hui...Li2019.7}, are found to be generally harder and more luminous than the other classes. This can be accounted for by the contribution of their intrabinary shocks in the X-ray production  \citep[see discussion in][]{Lee2018.864}.
 
For the progenitors of MSPs, namely the low-mass X-ray binaries (LMXBs), their formation rate per unit mass in GCs is known to be orders of magnitudes higher than that in the GF \citep[][]{Katz1975.199,Clark1975.199L}. This can be attributed to the frequent dynamical interactions in the central regions of GCs \citep[][]{Hui2010.714,Turk...Lorimer2013.436}. In some GCs, the stellar density can be high enough that multiple interactions of the binaries can occur \citep[][]{Verbunt2014.561}. With such complications, the evolution of compact binaries in GCs can possibly be different from those in the GF. And hence, it is not unreasonable to speculate that the characteristics of the MSPs, including the rotational, orbital and X-ray properties, in these two populations can be different.

Thanks to the sub-arcsecond angular resolution of {\it Chandra} X-ray Observatory, X-ray point sources can be resolved from the dense cores of GCs \citep[e.g.][]{Heinke2005.625,Bhattacharya2017.472,Bahramian2020.901,Oh2020.498} This enables us to identify the X-ray counterparts of MSPs by matching their radio timing positions with the X-ray source positions. With this sample, we can compare the X-ray properties of GC MSPs with their counterparts in the GF.

In this study, we first collected the updated samples of both X-ray and radio selected GC MSPs and normalized their X-ray properties. These allow us to convey a systematic analysis and compare their properties with those in the GF for investigating if there is any difference between these two populations of MSPs.

\section{Data Collection and Normalization}
To be consistent with \citet{Lee2018.864}, we define MSPs as the pulsars with rotational period $P<20$~ms in this work. 
With this criterion, a sample of 204 radio selected GC MSPs are collected from the online catalogue compiled by P. Freire.\footnote{\url{http://www.naic.edu/~pfreire/GCpsr.html}} On the other hand, we obtained an updated radio selected sample of 386 MSPs in the GF from the online catalogue maintained by West Virginia University. \footnote{\url{http://astro.phys.wvu.edu/GalacticMSPs/GalacticMSPs.txt}} The rotation period $P$ and the orbital period $P_{b}$ of the MSPs are also collected from these two catalogues. For obtaining the reliable estimates of spin-down rate $\dot{P}$ from a sub-sample of GC MSPs, please refer to Section 4.

To identify the X-ray counterparts of GC MSPs, we consider the sources detected by the Advanced CCD Imaging Spectrometer (ACIS) onboard {\it Chandra}. With its sub-arcsecond spatial resolution, ACIS enables the X-ray counterparts to be resolved from the crowded environment in GCs and provide their temporal and spectral information. The information of ACIS observations of MSP-hosting GCs are summarized in \autoref{tab:Table1}.

We found that 56 GC MSPs have their X-ray counterparts previously reported in the literature. Their properties and the relevant literature are summarized in \autoref{tab:Table2}. In order to compare the X-ray properties of GF MSPs reported by \cite{Lee2018.864}, we have to normalize the data with the same procedures adopted in their study. In the following, we describe the strategy for collecting the X-ray parameters in this work.

If a source has its X-ray spectrum characterized by an absorbed power-law (PL) model and with the spectral parameters reported in the existing literature, we adopt these reported properties in our work. However, since different studies have adopted different energy ranges in their X-ray analyses, it is necessary to normalize our X-ray fluxes of GC MSPs in the same energy band. 

With the aid of {\tt PIMMS}, we computed the absorption-corrected X-ray fluxes $f_x$ by integrating the spectral model in two energy ranges: 0.3-8 keV and 2-10~keV. While 0.3-8~keV is a standard band for analysing {\it Chandra} ACIS data, $f_x$ in 2-10~keV allow us to compare with those of GF MSPs reported by \cite{Lee2018.864}. Using the distances of the GCs $d$ (see \autoref{tab:Table2}), we computed the X-ray luminosities as $L_{x}=4\pi d^{2}f_{x}$.

This method allows us to obtain $L_{x}$ and the effective X-ray photon indices $\Gamma$ for the X-ray emitting MSPs in M13, M62, NGC 6397, Terzan5, and M22. Their parameters and the corresponding references are given in \autoref{tab:Table2}. 

For the sources that do not fulfill the criteria above, we analyzed the data directly by using CIAO (v.4.12). All the data were firstly reprocessed by using the {\tt chandra\_repro} script with updated calibration (CALDB v.4.9.2.1) and was filtered in the 0.3-8 keV energy band using {\tt dmcopy} task. All the data were reprocessed with subpixel event repositioning in order to facilitate a high angular resolution analysis. For the GCs with more than one observation, we first combined the data at different epochs by the {\tt merge\_obs} script. The images were subsequently produced with a binning factor of 0.5. By running a wavelet detection algorithm ({\tt wavdetect}) on the merged images with a range of scales (1.0, 1.414, 2.0, 2.828, 4.0), X-ray counterparts of the GC MSPs were identified if the sources were detected at a significance larger than 3$\sigma$ at the radio timing positions. 

The X-ray spectra of these counterparts were extracted by using {\tt specextract} in each individual observation. All the response files were generated by the same tool. The source extraction regions were selected so as to minimize the contamination of nearby sources. And the background regions were sampled in the circular source-free regions around the GCs with the radii in a range of 10-20 arcsec. All the spectral fittings in this work were performed in 0.3-8 keV with XSPEC (v.12.9). In view of low-counts data for most of the cases, all the analyses were performed with Cash statistics \citep{Cash1979ApJ.228.939C}, which enables us to perform fitting with unbinned data \citep[cf. Eq. 7 in ][]{Cash1979ApJ.228.939C}. This should give us less biased results than the binned analysis. If a source has been observed more than once (see \autoref{tab:Table1}), its spectra obtained from different observations were simultaneously fitted so as to obtain tighter constraints on its X-ray properties.  

In order to better constrain the spectral parameters and hence the X-ray fluxes, we took the column absorption $N_{H}$ as a fixed parameter throughout our analysis. If $N_{H}$ has been reported in the literature, the value is adopted for spectral fitting and computing the absorption-corrected flux. Otherwise, $N_{H}$ were estimated from the optical extinction $E(B-V)$ of the GCs \citep{Harris1996.112} through the correlation between these two quantities \citep{Guver2009.400}. 

 All the spectra were fitted with a simple absorbed PL model with XSPEC (i.e. TBABS $\times$ POWERLAW). With the multiplicative component CFLUX ( TBABS$\times$CFLUX$\times$POWERLAW), we obtained a robust estimate of the unabsorbed flux as well as its 1$\sigma$ uncertainty in both 0.3-8 keV and 2-10 keV energy bands.

With the aforementioned procedures, we have identified 56 confirmed X-ray detections of MSPs in 12 GCs and obtained the normalized estimates of their $L_{x}$ and $\Gamma$ (\autoref{tab:Table2}). The sample size is comparable with the X-ray selected MSPs found in the GF \citep[Tab. 1 in ][]{Lee2018.864}. 
The updated statistics of these radio and X-ray selected MSPs in different GCs are shown in the upper panel of \autoref{figure:figure1}. Following \cite{Lee2018.864}, we divided the MSPs into four different classes, isolated (Iso), redback (RB), black widow (BW) and  non-``spider" binaries (Oth). Their corresponding fractions of the radio/X-ray selected samples in the GF and GCs are shown in the lower panel of \autoref{figure:figure1}. 

Among our X-ray selected MSPs in 12 GCs, the samples in 7 GCs (i.e. M4, M13, M62, NGC 6397, Terzan5, M28, and M22) have also been covered by the catalogue compiled by \cite{Bahramian2020.901}. This allows us to cross-check the validity of our results. Within the tolerance of the statistical uncertainties, our estimates are found to be consistent with those given in \cite{Bahramian2020.901}. 

We found that 60 additional GC MSPs with known radio timing positions have been covered by the archival {\it Chandra} ACIS data serendipitously. This enables us to also search for their X-ray counterparts. Using the same procedures of data reduction and source detection described above, we did not find any additional X-ray emitting GC MSPs with detection significance larger than 3$\sigma$ in the merged images.

Despite the non-detections, the archival data still allow us to constrain the limiting $L_{x}$ of these 60 GC MSPs.
In examining the $L_{x}-\dot{E}$ relation for the MSPs in the GF, \cite{Lee2018.864} have shown that a less biased relation can be obtained from a survival analysis with the upper-limits of $L_{x}$ included (see Section 4). To obtain the $1\sigma$ limiting fluxes, we assumed a simple PL model with $\Gamma$=2 and $N_H$ adopted from the literature or inferred from the $E(B-V)$ of the corresponding GC. Together with the distances towards these GCs, we placed $1\sigma$ limiting luminosities of these additional 60 GC MSPs in $0.3-8$~keV and $2-10$~keV. The results are summarized in \autoref{tab:Table3}.

\section{Variability Analysis}
Apart from the periodic variations across the orbit resulted from different causes (e.g. intrabinary shock, eclipse of emission region, heating of companion surface), secular changes can also occur in a pulsar. The discoveries of RBs show that the properties of MSPs can vary considerably in different wavelengths as they switch between rotation-powered state and accretion-powered state \citep[e.g.][]{Papitto2013.501,Takata2014.785}. On the other hand, evidence of variable X-ray/$\gamma-$ray emission have also been observed from some isolated pulsars in GF \citep[e.g.][]{Lin2021.503,Takata2020.890,Hermsen2013Sci.339}. 

All these indicate that emission from a pulsar might not be as stable as previously thought. 
Since a number of GC MSPs in our sample have been observed by {\it Chandra} more than once, we are able to characterize their X-ray variabilities.

\cite{Bahramian2020.901} have included the results of variability test (i.e. $p-$values for Kolmogorov-Smirnov (K-S) test) for 7 out of 12 GCs in our sample. Among all 56 X-ray counterparts of GC MSPs in \autoref{tab:Table2}, 6 sources have $p-$values $<0.05$ for the K-S test as reported in \cite{Bahramian2020.901} (M62~B, NGC6397~A, M28~A, M28~I, M28~L and Terzan5~P). This indicates the possible variable X-ray emission from these sources. 

Our variability analysis was divided into into two parts: (1) long-term variability search and (2) short-term variability search. 
For (1), we searched for the possible X-ray flux variations of the targets among observations in different epochs. For (2), we searched for the possible variability within a single observation.  

For the long-term variability analysis, we only consider the observations in which the X-ray counterparts are detected with a significance $>3\sigma$ in a single exposure. In order to compare the $f_{x}$ of a given target in different epochs, we fitted its X-ray spectra obtained from individual observations. The response files generated in each observations can account for the possible instrumental variation among them. By fitting a simple absorbed PL model (with $N_{H}$ fixed) for each spectra, we obtained the estimates of absorption-corrected $f_{x}$ of a source in different epochs. Using these estimates, we constructed the long-term background-subtracted X-ray light curves for the subsequent analysis.

In order to identify the candidates that demonstrated long-term X-ray variability, we employed the Bayesian block algorithm that generates the optimal adaptive-width blocks \citep{Scargle2013ApJ.764.167S}. Even if the sequential data is not evenly sampled, the block-wise representation generated by this method can help to indicate local variability \citep[e.g.][]{Ahnen2016A&A.593A.91A}. Using the routine {\tt modeling.bayesian\_blocks.bayesian\_blocks} in the python library HEPSTATS, we have identified 10 sources require more than one block in modeling their long-term X-ray light curves. These are 47Tuc E, 47Tuc W, NGC6397 A, NGC6752 F, M28 A, M28 I, M28 L, Terzan5 A, Terzan5 P and Terzan5 ad. 

To scrutinize the significance of these variability candidates, we used two-sample Kuiper test \citep{Stephens1970.Kuiper} to compare their light curves with the uniform distributions determined by their corresponding mean fluxes. For this analysis, we utilized the routine {\tt kuiper\_twoside} in ASTROPY package (v.5.1). In this work, we consider a source has possible long-term X-ray variability if the $p-$value inferred from Kuiper test is $<0.05$. We found that only two sources from the short-list obtained from the Bayesian block analysis, M28 I and NGC6752 F, fulfill this criterion. Their long-term X-ray light curves are shown in \autoref{figure:figure2} with the identified Bayesian blocks illustrated. In the followings, we describe their temporal behaviors in further details.

M28~I (IGR J18245-2452), which is a RB has been found to swing between accretion-powered state and rotation-powered state \citep{Papitto2013.501}, is the most significantly variable X-ray source in our sample ($p\sim6.8\times10^{-24}$). This is consistent with the results reported by \cite{Linares2014.438} which has presented a detailed analysis of this source. 

The X-ray counterpart of the isolated MSP NGC6752~F can be detected in 6 out of 7 archival {\it Chandra} observations. The non-detection in the observation on 2017 July 25 (MJD 57959.83) can be ascribed to its relatively short exposure time ($\sim$18~ks). For this epoch, we placed a $1\sigma$ limiting $L_{x}$ of $3.9\times10^{30}$~erg/s and $1.9\times10^{30}$~erg/s in 0.3-8~keV and 2-10~keV respectively. In most of these observations, NGC6752~F behaves as a steady X-ray source except for the recent observation in 2017. In this epoch, its $L_{x}$ is found to increase by a factor of $\sim5$ in comparison with its previous level (\autoref{figure:figure2}). Kuiper test gives a $p-$value of 0.011 and suggests the variability can be significant. We further investigated whether such X-ray flux variation can be contaminated by the nearby bright sources. One MSP (NGC6752~D) and two cataclysmic variables (CVs) \citep[CX~1 \& CX~5 in][]{Forestell2014.441} are bright sources located at $\sim4.6$",$\sim7.4$", and $\sim6.5$" away from NGC6752~F respectively. There is no evidence of long-term X-ray variation found for NGC6752~D. Moreover, we do not find any resemblance between the long-term X-ray variation of NGC6752~F and its nearby CVs. Therefore, we concluded that the long-term X-ray variation of NGC6752~F is unlikely a result from the contamination of these bright sources.

We have also searched for the possible short-term variability within each observation windows by utilizing Gregory-Loredo variability algorithm \citep{Gregory.Loredo1992.398} for computing the odd ratios that the arrival times are not uniformly distributed in time. The algorithm is implemented in the CIAO tool {\tt glvary} which assigns a variability index according to the odd ratios. In this work, we set the criterion that a source demonstrates variability within a single observation if the inferred variability index is larger than 6. This implies the probability of this source to be variable is $\gtrsim90\%$. For avoiding the false alarm which results from the fluctuation due to low count statistics, we only consider the cases with more than 50 counts. 

Using the unbinned event lists and the corresponding effective areas, we have identified 3 GC MSPs, M28 I, NGC6397 A and Terzan5 P, which satisfy the aforementioned criterion in 6 observations. Their background-subtracted light curves are shown in \autoref{figure:figure3}. The binning of these light curves were determined by {\tt glvary} which give rise to the optimal variability.

Short-term X-ray variability of M28 I can be found in 3 observations of this cluster (Obs.ID: 9132, 9133 and 14616). All these observations were performed when M28 I were in the accretion-powered state. Its light curves on 2008 August 7 (Obs.ID: 9132) and 2008 August 10 (Obs.ID: 9133) show that the system was switching abruptly between the low state (with count rate of 0.02 count/s in 2-10 keV) and the high state ($>0.02$ count/s in 2-10 keV). These are consistent with the findings reported by \cite{Linares2014.438} which suggest this can be originated from the change of the magnetospheric radius due to the fluctuation of accretion flow. On the other hand, its light curve on 2013 April 28 (Obs.ID: 14616) also shows variable X-ray variation though it was less dramatic as the other two epochs. We note that this observation was close to the end of the thermonuclear outburst \citep{Linares2014.438} and just less than five days before it was found to switch back to rotation-powered state \citep{Papitto2013ATel.5069}.

For the RBs Terzan5~P and NGC6397~A, their X-ray flux variation can be originated from the intrabinary shocks. 
Short-term X-ray variability of NGC6397~A can be identified in a single observation on 2007 June 22 (Obs. ID: 7461). This observation started around the epoch when the pulsar was in superior conjunction (i.e. when the companion was located between the neutron star and the observer). The X-ray flux of the system began to rise gradually when the system was moving away from this phase. This is consistent with the findings reported by \cite{Bogdanov2010.709}.  A significant orbital X-ray modulation of Terzan5~P has been reported by \cite{Bogdanov2021.912} recently. Among all 18 {\it Chandra} ACIS observations of Terzan 5, short-term variability can be found in two observations on 2011 April 29 (Obs. ID: 13252) and 2011 September 8 (Obs. ID: 14339), which might be resulted from the fluctuations of the interactions between the pulsar wind and the stellar wind from the companion. 

We noted that the orbital variabilities of Terzan5~O, Terzan5~ad and M28~H were reported by \cite{Bogdanov2021.912,Bogdanov2011.730}, in which all the data were folded to their orbital periods. This is different from the scope of searching short-term variabilities within a single observation window in our current work. For these MSPs, their net counts are all less than 50 in all individual observations. Since they are lower than our predefined criterion for avoiding false alarm, they were not considered in our short-term variability search.

\section{Correlation \& Regression Analysis}
The spin-down power $\dot{E}$ of a pulsar is derived from $P$ and $\dot{P}$, $\dot{E}=4\pi^{2}I\dot{P}P^{-3}$, where $P$, $\dot{P}$, and $I$ are the rotational period, period derivative and the moment of inertia, respectively.
In examining GF MSPs, \cite{Lee2018.864} have found a $L_{x}-\dot{E}$ relation of  $L_{x}\simeq10^{31.05}\left(\dot{E}/10^{35}\right)^{1.31}$~erg~s$^{-1}$ in 2-10~keV. 
It will be instructive if one can construct a corresponding relation of the GC MSPs for comparison. 
Different from the GF MSPs, the MSPs in a GC are affected by the acceleration due to the cluster's gravitational potential. Hence, the Doppler effect can bias the measurements of $\dot{P}$ \citep[e.g.][]{Toscano1999MNRAS.307}. 
A large number of GC MSPs are found to have negative $\dot{P}$ \citep[cf.][]{Cheng2003.598}. 
This can complicate the estimation of the derived parameters such as $\dot{E}$ and surface magnetic field strength which is estimated as $B_{s}\simeq\sqrt{\frac{3c^{2}I}{2\pi^{2}R_{NS}^{6}}\dot{P}P}$, where $c$ is the speed of light and $R_{NS}$ is the radius of the neutron star which is assumed to be 10~km throughout this work.

\cite{Bogdanov2006.646} have adopted a King model to compute the cluster acceleration term for each MSP in 47 Tuc, and $\dot{E}$ are calculated by the intrinsic $\dot{P}$ which have the acceleration terms subtracted. Using these estimates, the authors examined the $L_{x}-\dot{E}$ relation in 47 Tuc and obtained $L_{x}\propto\dot{E}^{0.24\pm1.1}$. However, in examining the correlation between $L_{x}$ and $\dot{E}$ in their adopted sample by the Spearman rank correlation test, we found the correlation is very weak ($p$-value$\sim0.4$). The large uncertainties of their best-fitted $L_{x}-\dot{E}$ relation can be ascribed to this. While a King model provides a statistically reasonable model for the acceleration profile of a GC \citep{King1962.67.471K,Prager2017ApJ.845}, a small sample in this case can be hampered by the systematic uncertainties. 

For the GC MSPs in the binaries, long-term radio timing can provide another way for uplifting the contamination by their acceleration. \cite{Freire2017.471} have measured the time derivatives of orbital period $\dot{P}_{\rm b, obs}$ of 6 MSPs in 47 Tuc. Intrinsic $\dot{P}_{\rm int}$ can be estimated by $\dot{P}_{\rm int} = \dot{P}_{\rm obs} - \frac{\dot{P}_{\rm b, \rm obs}}{P_{\rm b}} P$. On the other hand, \cite{Prager2017ApJ.845} have also measured the $\dot{P_{\rm b}}$ for 9 MSPs in Terzan~5 as well as 47Tuc J which enable one to estimate their $\dot{P}_{\rm int}$. 

These studies of long-term radio-timing allow us to form a left-censored sub-sample of 16 MSPs with reliable estimates of $\dot{E}$ (12 X-ray detections + 4 upper-limits of $L_{x}$). Hereafter, we refer this sub-sample as Group A. Their $\dot{P}_{\rm int}$ as well as the derived $\dot{E}$ and $B_{s}$ are summarized in \autoref{tab:Table4}. 

Applying the Spearman rank test on the $L_{x}-\dot{E}$ relation on Group A, we obtain $p$-values of 0.013 and $7\times10^{-3}$ for $L_{x}$ in 0.3-8~keV and 2-10~keV respectively (\autoref{tab:Table5}). 
These suggest a much more significant correlation than the sample adopted by \cite{Bogdanov2006.646}. 

We have also examined the correlation with a larger data set by appending Group A with those have their King model corrected $\dot{P}$ reported in the literature. This can enlarge the sample size to 24 GC MSPs (20 confirmed X-ray detections and 4 upper-limits of $L_{x}$, see \autoref{tab:Table4}). We refer this sample as Group B hereafter. Spearman rank test on this group yields $p$-values of $10^{-3}$ and $2\times10^{-4}$ for $L_{x}-\dot{E}$ in 0.3-8~keV and 2-10~keV respectively (\autoref{tab:Table5}). Comparing with Group A, the improved significance of Group B is due to the enlarged sample. 

On the other hand, the $\dot{P}$ of M28~A ($\dot{P}=1.62\times10^{-18}$) is larger than those of the other GC MSPs by orders of magnitude. It is reasonable to argue that the effect of cluster acceleration on its $\dot{P}$ is not significant. Therefore, we further expanded Group B by including M28~A and we refer this sample as Group C hereafter. Spearman rank test on this group yields $p$-values of $2\times10^{-4}$ and $4\times10^{-5}$ for $L_{x}-\dot{E}$ in 0.3-8~keV and 2-10~keV respectively (\autoref{tab:Table5}) which indicate significant correlation between these two parameters. 

Apart from the improvement of the correlation, the uncertainties of $\dot{P}$ and hence $\dot{E}$ in our adopted sample are also reduced. The averaged percentage error of $\dot{P}$ for the 47 Tuc MSPs, as given by \cite{Freire2017.471} based on long-term pulsar timing, is $\sim65\%$. This is smaller than that in the sample adopted by \cite{Bogdanov2006.646} (i.e. $\sim75\%$). For the sample from Terzan 5, \cite{Prager2017ApJ.845} do not provide any error estimates for their $\dot{P}$. Since they are also obtained through long-term timing, we assumed the average percentage errors of the samples in \cite{Prager2017ApJ.845} are comparable with that in \cite{Freire2017.471} and computed the uncertainties accordingly. Combining with the other King model corrected $\dot{P}$, the overall averaged percentage error in Group B is found to be $68\%$. With M28~A included, the overall averaged percentage error in Group C becomes $65\%$.

The stronger correlation and the reduced uncertainties in our current sample prompt us to re-examine the $L_{x}-\dot{E}$ relation for GC MSPs and compare with their counterparts in the GF. In this work, all the regression analyses were performed in the framework of Bayesian inference. Instead of giving the point estimates, the posterior distributions of the parameters are reported so as to alleviate biases resulted from the small sample size. We adopt the {\tt R}-package {\tt LIRA} \citep{Sereno2016MNRAS.455.2149S} for our analyses, which not only allows an ordinary linear regression but also enables a survival regression analysis with the upper-limits of $L_{x}$ taken into account. In Bayesian framework, the conditional probability of a measurement $x$ with the given model $X$ is denoted by $P(x|X)$. In our analysis, $P(x|X)$ is proportional to a Gaussian for the detections. For the case that the observational results are expressed as upper limits, the conditional probability is truncated by a Heaviside function $H(x-x_{ul})$ where $x_{ul}$ is the upper limit for the left-censored data points \citep[cf. Appendix in][]{Willis2021MNRAS.503.5624W}. These treatments of the upper-limits are implemented in {\tt LIRA}. This can result in a less biased estimate for the $L_{x}-\dot{E}$ relation.

For the linear regression with a form of $\log L_{x}=\alpha\log\dot{E}+\beta$, the measurement uncertainties in both independent and dependent variables are taken into account. We assume a Student $t-$distribution and an uniform distribution as the priors for $\alpha$ and $\beta$ respectively. A 2D posterior probability distribution of these parameters was inferred through Markov chain Monte Carlo (MCMC). We have used four parallel chains with $2\times10^{6}$ iterations on each. The first $1\%$ of the samples from each chain were set as the initial burn-in. With these adaptation iterations excluded, all the other samples are used to approximate the posterior probability distribution.

Since our aim is to compare $L_{x}-\dot{E}$ relation between the MSPs in GC and GF, we analysed the data in both populations with the same procedures as aforementioned. For GC MSPs, we have started with the complete censored sample of Group C (i.e. with upper-limits of $L_{x}$ taken into account) as given in \autoref{tab:Table4}. For GF MSPs, we adopted the censored samples given by the Tables 1 and 2 in \cite{Lee2018.864}. 

The comparisons of the marginalized posterior distributions of $\alpha$ and $\beta$ inferred from Group C and the GF population are shown in \autoref{figure:figure4}. For GF MSPs, the distribution of the slope (i.e. $\alpha$) is found to be peaked around $\sim0.9$ and $\sim1.2$ in 0.3-8~keV and 2-10~keV respectively. The latter one is consistent with the point estimate reported by \cite{Lee2018.864} in the same energy band. On the other hand, the posterior distribution of $\alpha$ inferred from the GC MSPs in Group C is peaked around $\sim0.6$ and $\sim0.8$ in 0.3-8~keV and 2-10~keV respectively. The comparison of the $L_{x}-\dot{E}$ relation in these two populations suggests a possible difference.  

The plots the $L_{x}-\dot{E}$ relations in \autoref{figure:figure4} are shown for a further comparison. Uncertainties estimated from the ranges centered at the peaks and bracket $68\%$ of the samples in the marginalized posterior distributions are illustrated as the shaded regions in these plots. The difference between these two populations is also suggested by the lack of overlap between their shaded regions.

By visually examining the plots of $L_{x}-\dot{E}$ for the GC MSPs in \autoref{figure:figure4}, the asymmetric distribution of the data points above and below the best-fit relation suggests the fitting is far from desirable. We speculate that M28~A can possibly be an outlier. Since the regression analysis is weighed by the reciprocal of the uncertainties of the data, the fact that the errors of $\dot{E}$ and $L_{x}$ of M28~A is much smaller than those of the other GC MSPs can result in a strong bias towards this single data point.

To quantify this issue, we have computed the interquartile range (IQR) of $\log L_{x}$. For detecting outliers, we adopted the conventional criterion of 1.5 times of IQR, which is found to be $\log L_{x}=29.96-31.72$ (0.3-8~keV) and $\log L_{x}=28.92-31.61$ (2-10~keV) in Group C. With this procedure, M28~A is the only source lies outside 1.5$\times$IQR of Group C.

Distinctions between M28~A and the majority of GC MSPs can also be discussed in terms of physical reasons. First, its characteristic age ($\tau\sim3\times10^{7}$~yrs) is much smaller than the other MSPs. Apart from its high value of $L_{x}$, the X-ray pulses of this isolated MSP have a very narrow profile which suggests its non-thermal nature with the origin from the magnetospheric accelerator \citep{Du2015ApJ.801.131D}. Also, its X-ray emission can be detected at energies up to $\sim50$~keV. All these make the X-ray properties of M28~A very different from the thermal X-rays originated from most of the isolated MSPs in GCs. Furthermore, it is one of the two MSPs which have glitches detected so far \citep{Cognard2004ApJ.612L.125C,McKee2016MNRAS.461.2809M}. This might suggest M28~A is more similar to  young energetic pulsars than a typical MSP.

To be consistent, although the best-fit $L_{x}-\dot{E}$ relation for the GF MSPs in \autoref{figure:figure4} is reasonable, we have also searched for the possible outliers in the GF sample with the same procedure we applied in the GC sample. In 0.3-8~keV, PSR~J0218+4232 and PSR~B1937+21 are found lying outside $1.5\times$IQR of $\log L_{x}$ ($28.79-32.72$) of the GF sample. 
And in 2-10~keV, only PSR~J0218+4232 lies outside the corresponding range ($\log L_{x}=27.32-33.19$). This prompts us to re-do the fitting by considering both of them as the outliers.

With the outliers removed from both GC and GF samples, we have re-run the regression analysis for inferring $L_{x}-\dot{E}$~relation. The results are shown in \autoref{figure:figure5}. For the GF MSPs, excluding PSR~J0218+4232 and PSR~B1937+21 only results in a slightly flatter $L_{x}-\dot{E}$~relation in comparison with the case including all the samples (i.e. \autoref{figure:figure4}). Their difference can be reconciled with the tolerance of their uncertainties. 

On the other hand, in the case of GC MSPs, the posterior distribution of $\alpha$ inferred with M28~A removed (i.e. Group B) is peaked around $\sim0.4$ and $\sim0.5$ in 0.3-8~keV and 2-10~keV respectively. The relation appears to be much flatter than that shown in \autoref{figure:figure4}. And we found that the quality of the fitting is much improved as similar number of data points are above and below the best-fit line. With the outliers excluded, the comparison between the posterior distributions of $\alpha$ and $\beta$ suggests the difference in the $L_{x}-\dot{E}$ relation between GC MSPs and GF MSPs becomes more significant.

A recent study of the MSPs in Terzan~5 has suggested a positive correlation between $L_{x}$ and the X-ray hardness \citep[cf. Figure 3 in][]{Bogdanov2021.912}. With the effective photon index given in \autoref{tab:Table2} as a measure of X-ray hardness of GC MSPs (i.e. smaller $\Gamma$ implies harder X-ray emission), we are able to examine if this relation can be found in the full sample of X-ray selected GC MSPs with the photoelectric absorption corrected. Spearman rank test suggests a strong correlation between $L_{x}$ and $\Gamma$ with a $p-$value of $5.6 \times 10^{-16}$ and $5.3 \times 10^{-5}$ in both 2-10 keV and 0.3-8 keV. For the non-detections in \autoref{tab:Table3}, the upper-limits of $L_{x}$ are calculated by assuming a PL model with fixed $\Gamma$ and hence they are not very informative for examining the relation between $L_{x}$ and $\Gamma$. Therefore, we ignored the upper limits in the regression analysis of $\log L_{x}-\Gamma$.

Using the procedures as described above, we obtained the marginalized posterior distributions for the parameters $a$ and $b$ in the assumed linear relation of $\log L_{x}=a\Gamma + b$. We have applied the same analysis on the GF sample as given by \cite{Lee2018.864}. The comparison of this relation between these two populations are shown in \autoref{figure:figure6}. No significant difference in terms of X-ray luminosity and hardness is found between the MSPs in GCs and GF. 

From the literature, we have also obtained the information of whether the X-rays from the GC MSPs are dominated by thermal or non-thermal emission which are summarized in \autoref{tab:Table2}. In \autoref{figure:figure6}, we differentiate the non-thermal dominant and thermal dominant cases by different symbols. We found that the non-thermal dominant X-ray GC MSPs are characterized with an effective photon index of $\Gamma < 2$ in our analysis. On the other hand, the thermal X-ray emitters are generally characterized with $\Gamma > 2$. 
The fact that the non-thermal dominant X-ray GC MSPs are generally more luminous can be due to the presence of additional harder X-ray components from the intrabinary shock in these systems \citep{Lee2018.864}. 

Many studies have shown that the final state of a MSP strongly depends on the initial mass of its companion and the orbital separation \citep[e.g.][]{Tauris2011ASPC.447.285T,Liu.Chen2011MNRAS.416}. Furthermore, evolutionary status of the companion at the onset of the Roche lobe overflow (RLO) is suggested to be a key factor in determining the timescale of the mass transfer phase which can directly affect the nature of the MSP \citep{Tauris2011ASPC.447.285T,Tauris1999A&A.350}. The longer the mass transfer phase (i.e. more mass accreted by the neutron star) will result in a faster rotating MSP \citep[cf. Fig. 5 in][]{Liu.Chen2011MNRAS.416}.

If the orbital separation of the progenitor is wider, the companion needs to be more evolved by the time it fills its Roche lobe and transfers its mass to the neutron star. This will lead to a shorter mass transfer phase and hence a relatively slower rotating MSP. Hence, This suggests that investigating the correlation between $P_{b}$ and $P$ (i.e. Corbet diagram) can provide a fossil record for the evolutionary history of compact binaries. 

The distributions of $P$ vs $P_{b}$ for radio/X-ray selected samples in GF and GCs are shown in \autoref{figure:figure7}. We started the analysis with all the MSP binaries. With Spearman rank test, we found that $P$ and $P_{b}$ in radio selected sample of GF MSPs are strongly correlated ($p$-value$\sim7\times10^{-9}$). On the other hand, the correlation becomes weaker in their X-ray selected sample and it is marginally significant ($p$-value$\sim0.06$), which can be due to the much reduced sample size of the X-ray emitting GF MSPs. 

However, in both radio selected and X-ray selected samples of GC MSPs, we do not find any evidence for the correlation between $P$ and $P_{b}$  ($p-$value $>0.1$ in both cases). The comparisons of $P-P_{b}$ correlation test for the MSPs in GF and GCs are summarized in \autoref{tab:Table6}. The lack of such correlation in GC MSPs is likely a results of dynamical interactions (see the discussion in Section 6). 

For the GF MSPs, we further examined $P-P_{b}$ correlation from each types of MSP binaries. In \autoref{figure:figure7}, the binaries with different nature of companion are represented by different symbols. By running the Spearman rank correlation test on each types of MSP binaries, we found that only those with a helium white dwarf (He WD) as the companion show a significant correlation between $P$ and $P_{b}$ ($p-$value = $7.0\times10^{-4}$). This can be accounted by the relatively simple evolutionary track with wide LMXBs as progenitors \citep[Case B RLO][]{Tauris2011ASPC.447.285T}. For the other types of MSP binaries (e.g. spider MSPs, MSPs with CO WD companion), the lack of $P-P_{b}$ correlation might be a result of the more complex evolutionary channels for their formation \citep{Tauris1999A&A.350,Tauris2011ASPC.447.285T}. 

The $P-P_{b}$ correlation found in radio selected GF MSPs leads us to perform the Bayesian regression analysis by assuming a linear relation of 
$\log P=m \log P_{b}+c$. We have run the analysis for the cases with the full sample as well as only with those have a He WD companion. The marginalized posterior probability distributions of $m$ and $c$ as well as the best-fit relation are shown in \autoref{figure:figure7}.

\section{Globular cluster MSPs vs. Galactic field MSPs}
To investigate whether the frequent stellar interactions in GCs have any effects on the physical properties of their MSPs, we compare a set of parameters of GC MSPs with those of their counterparts in the GF through standard statistical tests. Six parameters, including $\Gamma$, $L_{x}^{2-10}$, $P_{b}$, $P$, $B_{s}$ and $\dot{E}$, are chosen in this analysis.  

Before any comparison, we first constructed the unbinned empirical cumulative distribution functions (eCDFs) for each parameter (\autoref{figure:figure8}-\ref{figure:figure10}). To quantify the difference between any two eCDFs in consideration, we employed two different non-parametric statistical tests: two-sample Anderson-Darling (A-D) test and Kolmogorov-Smirnov (K-S) test. While K-S test is widely used in literature, we notice it has several drawbacks.\footnote{https://asaip.psu.edu/articles/beware-the-kolmogorov-smirnov-test/}. For example, it is not sensitive to supremum distance between two eCDFs far away from their centers. On the other hand, two-sample A-D test provides a more sensitive method in identifying the difference between two distributions. $p-$values inferred from both A-D and K-S tests in different comparisons are listed in \autoref{tab:Table7} and \ref{tab:Table8}. In this study, if the $p-$value inferred from either test is $\lesssim0.05$, the difference between the eCDFs is considered to be plausible and will be further discussed. 

We first compared the X-ray luminosities and hardness between the MSPs in GF and GCs. The eCDFs of $L_{x}^{2-10}$ and $\Gamma$ for all known X-ray emitting MSPs in these two populations are shown in \autoref{figure:figure8}. While we do not find any significant difference in $\Gamma$ between the X-ray selected samples in GF and GCs ($p$-value $>0.1$ in both A-D and K-S test), A-D test suggests a marginal difference in $L_{x}^{2-10}$ ($p$-value $\sim0.02$). In examining their eCDFs, the possible difference can be in the low luminosity range of 
$L_{x}^{2-10}\lesssim5\times10^{29}$~erg/s. This can be due to the fact that there are more nearby systems in the GF which allow fainter MSPs to be detected. 

We have also divided the full sample of X-ray selected MSPs into four classes (Iso, RB, BW and Oth) and compared the corresponding classes in GCs and GF (e.g. RBs in GCs vs RBs in GF). The results are shown in \autoref{figure:figure9} and \autoref{tab:Table7}. We found the $p-$values inferred from both tests are $>0.05$ for all cases and hence no significant difference in the X-properties between the corresponding classes in GCs and GF can be identified with the current sample. 

For comparing $P_{b}$, $P$, $B_{s}$ and $\dot{E}$, we have examined both radio selected and X-ray selected samples in order to investigate the possible selection effect imposed by X-ray observations. For the comparison of $B_{s}$ and $\dot{E}$, we started with the outliers in both populations (i.e. M28~A, PSR J0128+4232 and PSR~B1937+21) excluded.

We found that all these four parameters are significantly different between the radio selected MSPs in GCs and GF (see the first row of \autoref{figure:figure10}). The most obvious differences are found between their $P_{b}$ and $\dot{E}$, in which A-D test gives $p-$values of $\sim10^{-5}$ and $10^{-4}$ respectively (see \autoref{tab:Table8}).

However, when we compare the X-ray selected samples of these two populations, the differences in their $P_{b}$ and $\dot{E}$ distributions disappear (see the second row of \autoref{figure:figure10}). For example, the significance for the differences between their $P_{b}$ and $\dot{E}$ drop drastically ($p-$value $\sim0.7$ from A-D test, see \autoref{tab:Table8}). This clearly indicates the presence of selection effect imposed by X-ray detections. 

To further examine such effect, we tabulate the medians of these four parameters of both X-ray selected and radio selected samples in GF and GCs (see \autoref{tab:Table9}). We have also compared the X-ray/radio selected eCDFs in GF and GCs, which are shown in the third and the forth rows in \autoref{figure:figure10} respectively. 

In view of the large uncertainties of $B_{s}$ and $\dot{E}$ for the GC MSPs, we examined the possible impact of the measurement errors on the aforementioned inference by Monte Carlo sampling. We assumed a Gaussian distribution centered on each of the observed values of $\log B_{s}$ and $\log\dot{E}$ in \autoref{tab:Table4} with the corresponding errors as the standard deviations. A set of simulated sample can then be randomly drawn from each of these distributions. In total, 10000 sets of simulated samples were generated in our experiment. For each set of sample, we have run A-D test to compare its eCDFs with those of GF MSPs and computed the corresponding $p-$values. 

In \autoref{figure:figure11}, we show the empirical distributions of $p-$values obtained from the aforementioned Monte Carlo method. The green dashed lines illustrated the $p-$values computed with the observed data (cf. \autoref{tab:Table8}). Taking $p<0.05$ as the benchmark for two distributions being different, we estimated the probabilities of obtaining $p<0.05$ from these empirical distributions. For comparing $\dot{E}$ between the radio selected MSPs, 100\% of our simulated data result in $p<0.05$. On the contrary, none of the simulated data leads to $p<0.05$ in comparing $\dot{E}$ between the X-ray selected MSPs. These results support our assertion that $\dot{E}$ of the MSPs in GC and GF are different in the radio selected samples but such difference is diminished in the X-ray selected samples. For comparing $B_{s}$ between the MSPs between GCs and GF, we found that $\sim99\%$ and $\sim96\%$ of the simulated data give $p-$value below the benchmark in comparing the radio selected and X-ray selected samples respectively. These support the conclusion that the distributions of $B_{s}$ for the MSPs in GCs and GF are different, regardless of X-ray selected or radio selected.

We have repeated the analysis of comparing $B_{s}$ and $\dot{E}$ between the MSPs in GCs and GF with the outliers included. The comparisons of eCDFs and the empirical distributions of $p-$values obtained from the Monte Carlo method are shown in \autoref{figure:figure12} and \autoref{figure:figure13} respectively. We found that the results are fully consistent with those inferred from the analysis with the outliers removed. In \autoref{figure:figure13}, while 100\% of our simulated data result in $p<0.05$ for the comparison of $\dot{E}$ between the radio selected MSPs in GCs and GF, there is only $0.1\%$ of simulated data below this benchmark in comparing the same parameter between the X-ray selected MSPs in these two populations. On the other hand, in comparing $B_{s}$ between the MSPs between GCs and GF, we found that $100\%$ and $98\%$ of the simulated data show $p<0.05$ in the radio selected and X-ray selected MSPs respectively.

For the GF MSPs, both $P_{b}$ and $P$ in their X-ray selected sample are significantly shorter than those in their radio selected sample (\autoref{tab:Table9}). A-D test yields the $p-$values of $\sim10^{-3}$ in comparing the corresponding eCDFs which indicates such differences are significant (see the third row of \autoref{figure:figure10} and \autoref{tab:Table8}). On the other hand, we found that the surface magnetic field strength $B_{s}$ of both radio selected and X-ray selected GF MSPs are very similar (\autoref{figure:figure10} and \autoref{tab:Table8}). Since $\dot{E}$ scales as $\dot{E}\propto B_{s}^{2}P^{-4}$, the difference of this parameter between the radio selected and X-ray selected GF MSPs is expected as A-D test yields a $p-$value of $\sim10^{-4}$. 

It is clear that the X-ray observations have detected MSPs in the GF with faster rotation (i.e. small $P$) and hence more powerful (i.e. higher $\dot{E}$) (see \autoref{tab:Table9}).

For the X-ray emitting GC MSPs, however, we do not find any significant selection effect imposed by X-ray observations in the GC MSP population (see \autoref{tab:Table9}, \autoref{figure:figure10}/\autoref{figure:figure12}). For all four parameters considered in this analysis (see the fourth row of \autoref{figure:figure10}/\autoref{figure:figure12}), neither A-D test nor K-S test can identify any significant difference between the radio selected and X-ray selected samples in GC MSPs (see \autoref{tab:Table8}). For example, different from the case of GF MSPs, the X-ray selected MSPs in GCs do not appear to rotate significantly faster than their radio selected sample ($p\sim0.13$ by A-D test). 

There is another interesting feature found in comparing these two populations. Regardless of whether it is X-ray or radio selected, GC MSPs generally rotate slower than those in GF. For the X-ray emitting MSPs, since their $\dot{E}$ are comparable in GCs and GF, the slower rotating GC MSPs suggests their surface magnetic field $B_{s}$ should be stronger. Such expected difference can be seen by comparing their eCDFs and medians. With the outliers excluded in both GCs (i.e. M28~A) and GF (i.e. PSR~J2018+4232 and PSR~B1937+21), a difference between the X-ray selected MSPs in GCs and GF is suggested by both A-D and K-S tests ($p\sim0.02$). A more significant difference of $B_{s}$ between the radio selected MSPs in GCs and GF are indicated by both tests ($p\lesssim5\times10^{-3}$). The conclusions are unaltered when the outliers are included in the comparison (\autoref{tab:Table8} \& \autoref{tab:Table9}).

\section{Summary and Discussion}
We have performed a systematic analysis of the rotational, orbital and X-ray properties of MSPs in GCs and compared with those in the GF. The major results are summarized as follows:

\begin{enumerate}
    \item GC MSPs generally rotate slower than those in the GF.
	\item While X-ray observations tend to pick the MSPs with faster rotation in the GF, we do not find such selection effect in the GC MSP population.
	\item Surface magnetic field ($B_{s}$) of GC MSPs are apparently stronger than those in the GF.
	\item For the MSP binaries, strong correlation is found between the rotation period and the orbital period in the GF population. However, such correlation is absent in the GC MSP binaries.
	\item Although the distributions of X-ray luminosity ($L_{x}$) and hardness ($\Gamma$) for the MSPs in GCs are comparable with those in the GF, the GC MSPs apparently follow a different $L_{x}-\dot{E}$ relation.
\end{enumerate}

All these findings suggest that dynamical interactions in GCs can alter the evolution of MSPs/their progenitors and leave an imprint on their X-ray emission properties. Here we discuss the implications of our results. 

One most distinguishable properties between the radio selected MSPs in GCs and GF is their distributions of $P_{b}$ (\autoref{figure:figure10}). It is clear that there is a lack of wide orbit MSP binaries in GCs. This can be accounted by the frequent stellar encounters in GCs. Numerical studies have shown that close encounters between stars and binaries can affect the orbital parameters and dramatically alter the evolution of the binaries \cite[][]{Benacquista2013.16}. If the initial binding energy of a primordial binary is larger than the average kinetic energy of the neighboring stars in the cluster, the encounter can lead to orbital shrinkage with the orbital binding energy transferred to the neighboring stars \cite[][]{Heggie1975.173}. 

It is instructive to compare the average orbital binding energy $\langle E_{b}\rangle$ of the radio selected GC MSPs with the averaged kinetic energy of the neighboring stars $\langle E_{*}\rangle$ in their hosting clusters. For each MSP binary, we computed $E_{b}$ by $GM_{\rm psr}M_{c}/2a$ where $M_{\rm psr}$, $M_{c}$ and $a$ are the mass of pulsar, the mass of companion and the semi-major axis of the orbit respectively. We fixed $M_{\rm psr}$ at 1.35$M_{\odot}$ for all systems. Both $a$ and $M_{c}$ are taken from the ATNF pulsar catalog \citep{Manchester2005.129} by assuming an orbital inclination of $i=60^{\circ}$. With these estimates, the average orbital energy of the radio selected GC MSP binaries is found to be $\langle E_{b}\rangle\sim2.3\times10^{45}$~ergs. On the other hand, the corresponding value of the radio selected GF MSP binaries is $\langle E_{b}\rangle\sim8.5\times10^{44}$~ergs which is about three times lower.  

For $\langle E_{*}\rangle$, we calculated by averaging the characteristic value for each MSP-hosting GC. We computed $E_{*}$ by $\frac{1}{2}M_{*}\sigma^{2}_{*}$, where $M_{*}$ and $\sigma_{*}$ are typical mass and the velocity dispersion of the neighboring stars in a GC. Values of $\sigma_{*}$ were adopted from  \cite{Harris2010arXiv1012.3224H}. For estimating $M_{*}$, we took the mass of a main sequence corresponding to the spectral type of the integrated cluster light for each GC given in \cite{Harris2010arXiv1012.3224H}. It is interesting to note that $\langle E_{*}\rangle\sim1.6\times10^{45}$~ergs is rather close to the estimate of $\langle E_{b}\rangle$ for the GC MSP binaries. The similarity of these two quantities might indicate the past interactions between MSP binaries (and/or their progenitors) and the neighboring stars in the cluster, which can lead to equipartition among orbital energy, recoil kinetic energy of the binaries and the kinetic energy of the stars in GCs.

On the other hand, for any primordial binaries with wide orbits which have initial binding energy smaller than the average kinetic energy of the neighboring stars in the cluster, they are prone to be destroyed through the single-binary interaction \cite[][]{Heggie1975.173,Benacquista2013.16}. All these can make the recycling process more complicated than their counterparts in the GF. This might explain the absence of $P_{b}-P$ correlation for the MSP binaries in GCs. 

Disturbance on the recycling process for the MSPs in GCs might also account for their slower rotations. Since mass transfer can be disrupted by the agitation of frequent stellar encounter \citep{Verbunt2014.561}, this can leave the MSP with intermediate rotation period $P$ which is consistent with our results. 

We notice that \cite{Konar2010.409} have reported an opposite conclusion (i.e. GC MSPs rotate {\it faster} than their counterparts in GF). The difference between their results and ours can be accounted by the difference in the adopted samples. While we selected the MSPs with the criterion $P<20$~ms, \cite{Konar2010.409} selected their sample with $P<30$~ms. And most importantly, the sample size in our study is $\sim3$ times larger than that adopted by \cite{Konar2010.409}. And we have more fast rotating MSPs in our sample. The average $P$ of MSP in the GF/GCs are 7.75/5.70 ms in their sample \citep[cf. Tab. 1 in][]{Konar2010.409}. The corresponding values in our radio selected sample are found to be 3.73/4.34 ms (\autoref{tab:Table9}). On the other hand, with a much larger sample, our results confirm the scenario suggested by \cite{Verbunt2014.561}.

We have also examined whether such difference can be a result of observational effect. Since GC MSPs are generally located further than their GF counterparts, detecting fast rotating pulsars in GCs by radio observations can be more difficult because of the possible broadening of their pulses by scattering. This can possibly result in a MSP population in GCs with slower rotation than that in the GF. This prompts us to compare the pulse width between these two populations by taking the estimates of the pulse widths at 50\% of their peaks in ATNF catalog \citep{Manchester2005.129}. Both A-D and K-S tests yield a $p$-value $>0.1$ and hence there is no indication that the radio pulses of GC MSPs are broader.

Another evidence against the aforementioned hypothesis is the detection of the fastest known pulsars in Terzan~5, namely Terzan5~ad ($P\sim$1.4 ms). Despite the fact that the pulsars in Terzan~5 have the highest dispersion measure among all MSPs, the discovery of Terzan5~ad shows that the improvement in instrumentation and search techniques in the radio surveys have greatly overcome the bias in detecting fast pulsars in GCs. For example, an effective temporal resolution of $\sim0.3$~ms was achieved in the pulsar search towards Terzan 5 \citep{Ransom2005Sci.307.892R,Hessels2006Sci.311.1901H}. Together with the fact that no known bias against the detection of slow pulsars, we do not find any convincing argument that the difference between the rotational period distributions of these two populations is a result of observational bias. Hence, we conclude the result that GC MSPs generally rotate slower than GF MSPs is intrinsic.

\begin{figure}
\centering
\includegraphics[width=6in, height=3.5in, angle=0]{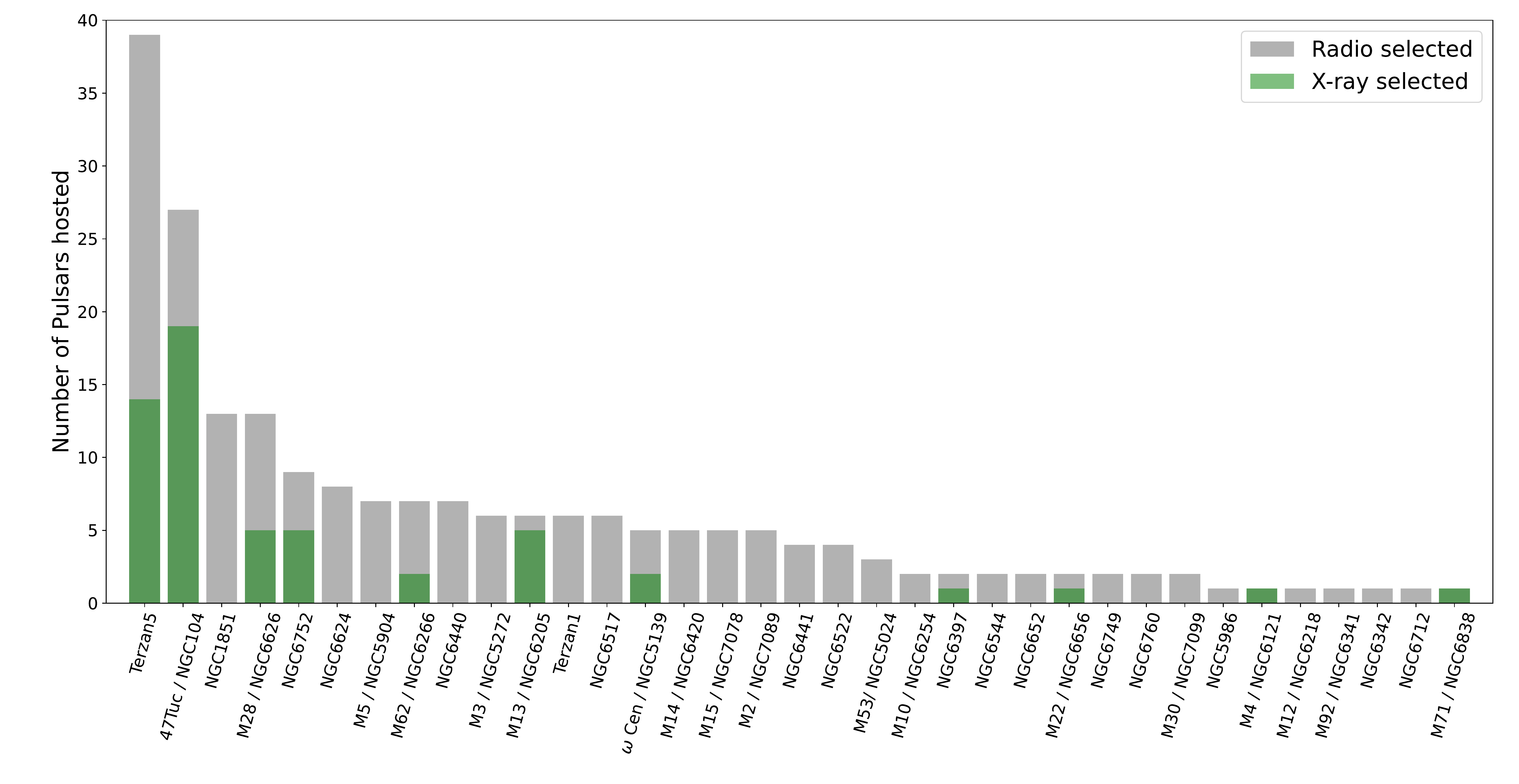}
\includegraphics[width=4in, height=4in, angle=0]{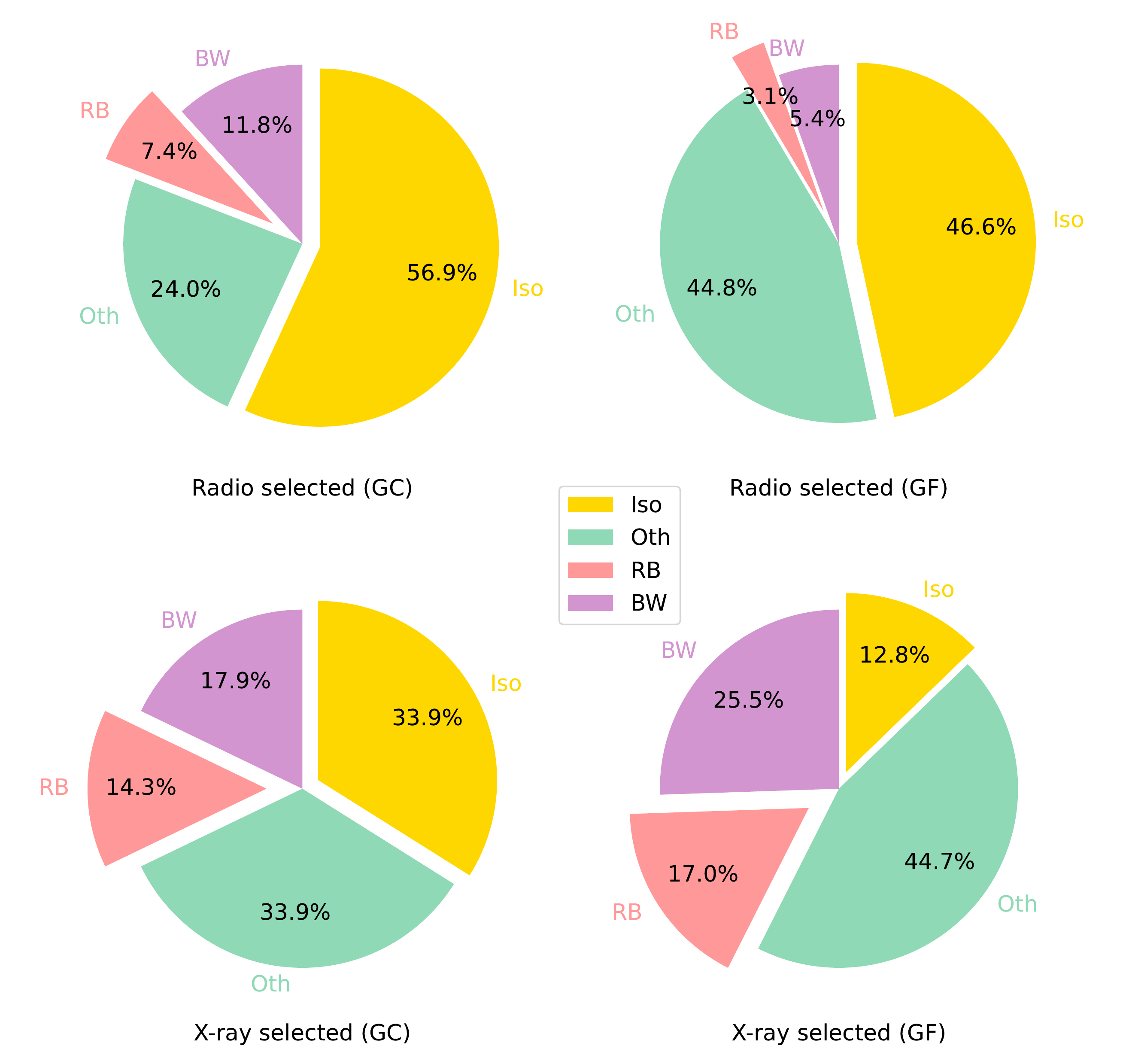}
\caption{\label{figure:figure1} ({\it Upper panel}) Updated statistics of X-ray selected and radio selected GC MSPs adopted in this work. ({\it Lower panel}) The distributions of different classes of the known radio/X-ray selected MSPs in GCs and GF.}
\end{figure}

\begin{figure}
\centering
\includegraphics[width=6in, height=2in, angle=0]{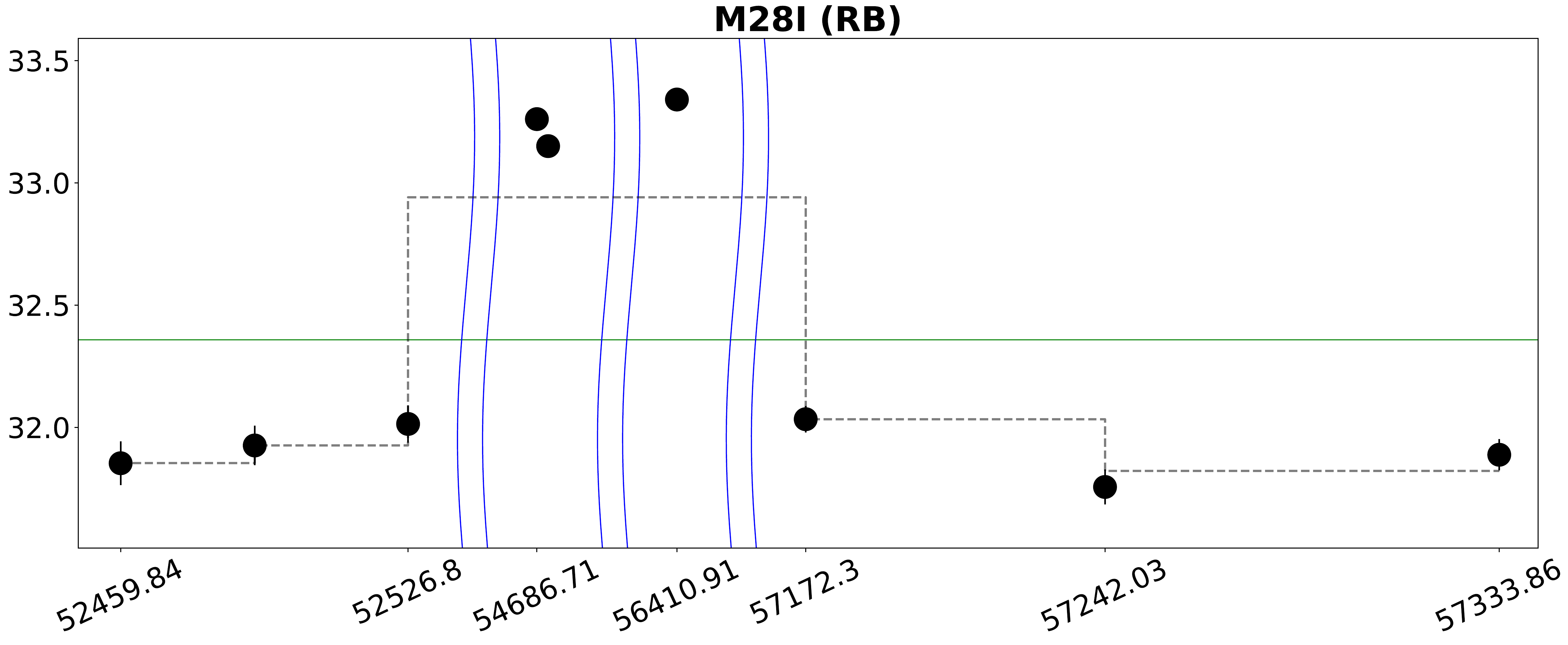}
\includegraphics[width=6in, height=2in, angle=0]{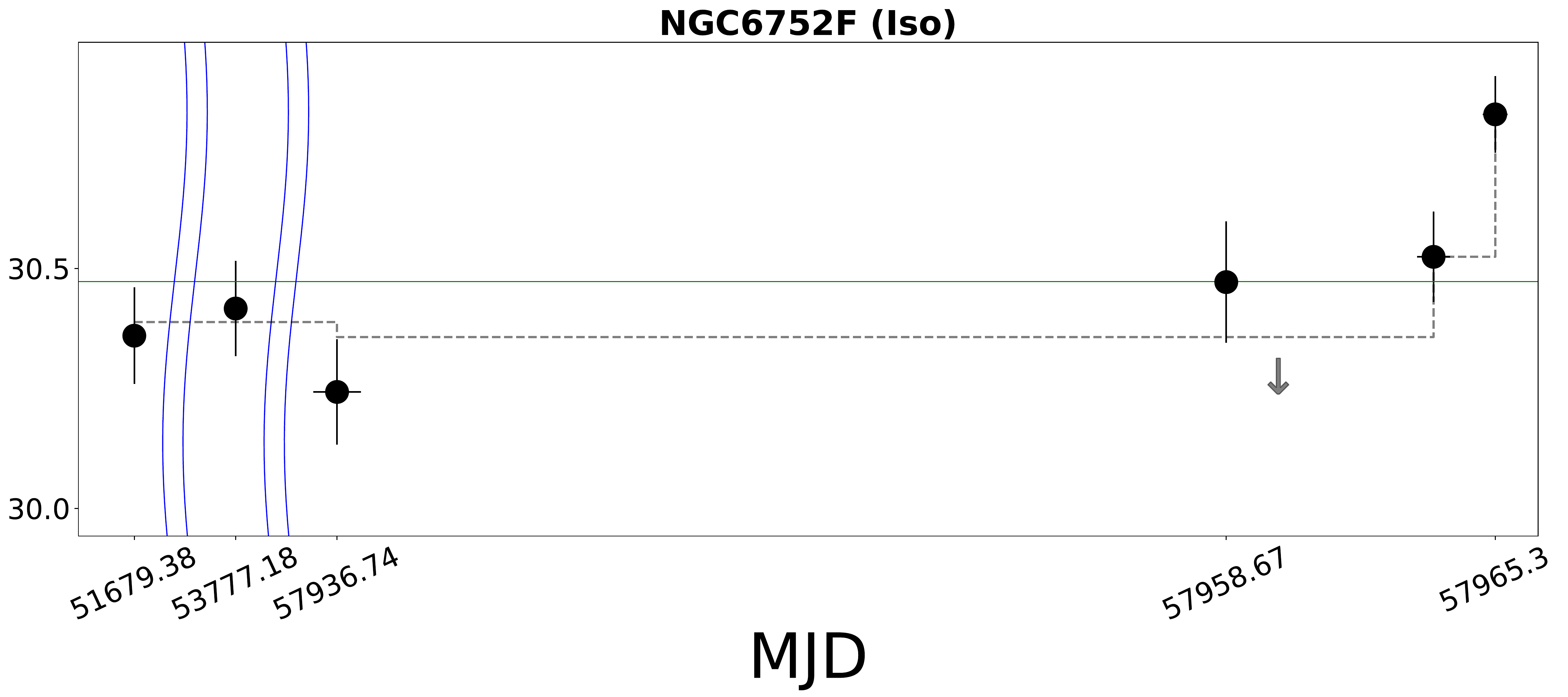}
\caption{\label{figure:figure2} X-ray light curves of the GC MSPs which exhibit long-term flux variability. The Bayesian Block representation of the light curves are illustrated by the dashed lines. The green horizontal lines represent the levels of the mean flux in each case. The time axes are in linear scale with the gaps larger than 1 years intermitted by the blue stripes. The labeled epochs correspond to the midpoints of the observations. The arrow in the plot of NGC6752~F represent the 1$\sigma$ upper limits of $L_x$ for that particular epoch.}
\end{figure}

\begin{figure}
\centering
\includegraphics[width=6in, angle=0]{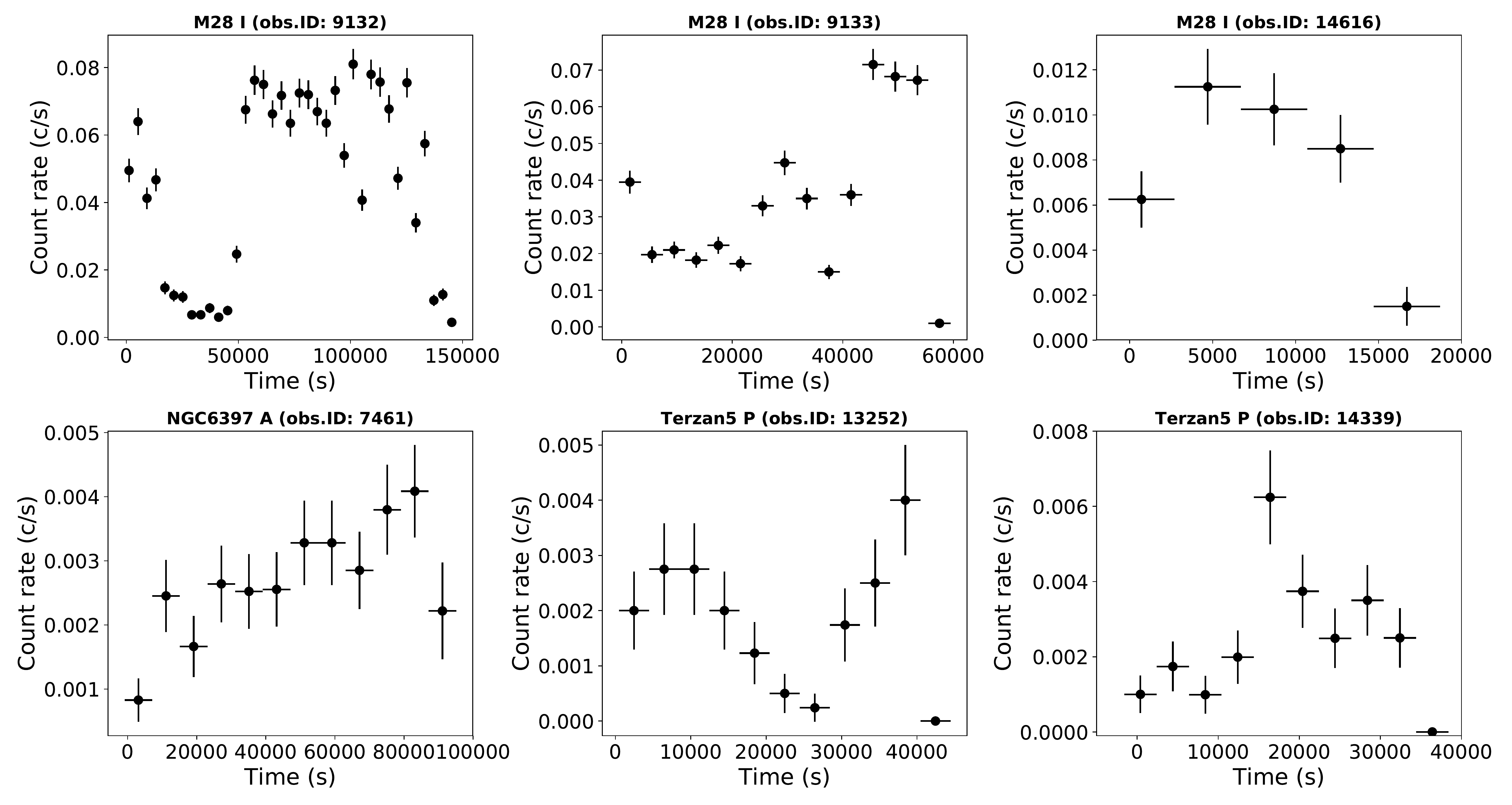}
\caption{\label{figure:figure3}  Background-subtracted X-ray light curves of GC MSPs from the individual observations in which short-term variabilities are identified. The optimal binning are determined by CIAO tool {\tt glvary}.}
\end{figure}

\begin{figure}
\centering
\includegraphics[width=7in, angle=0]{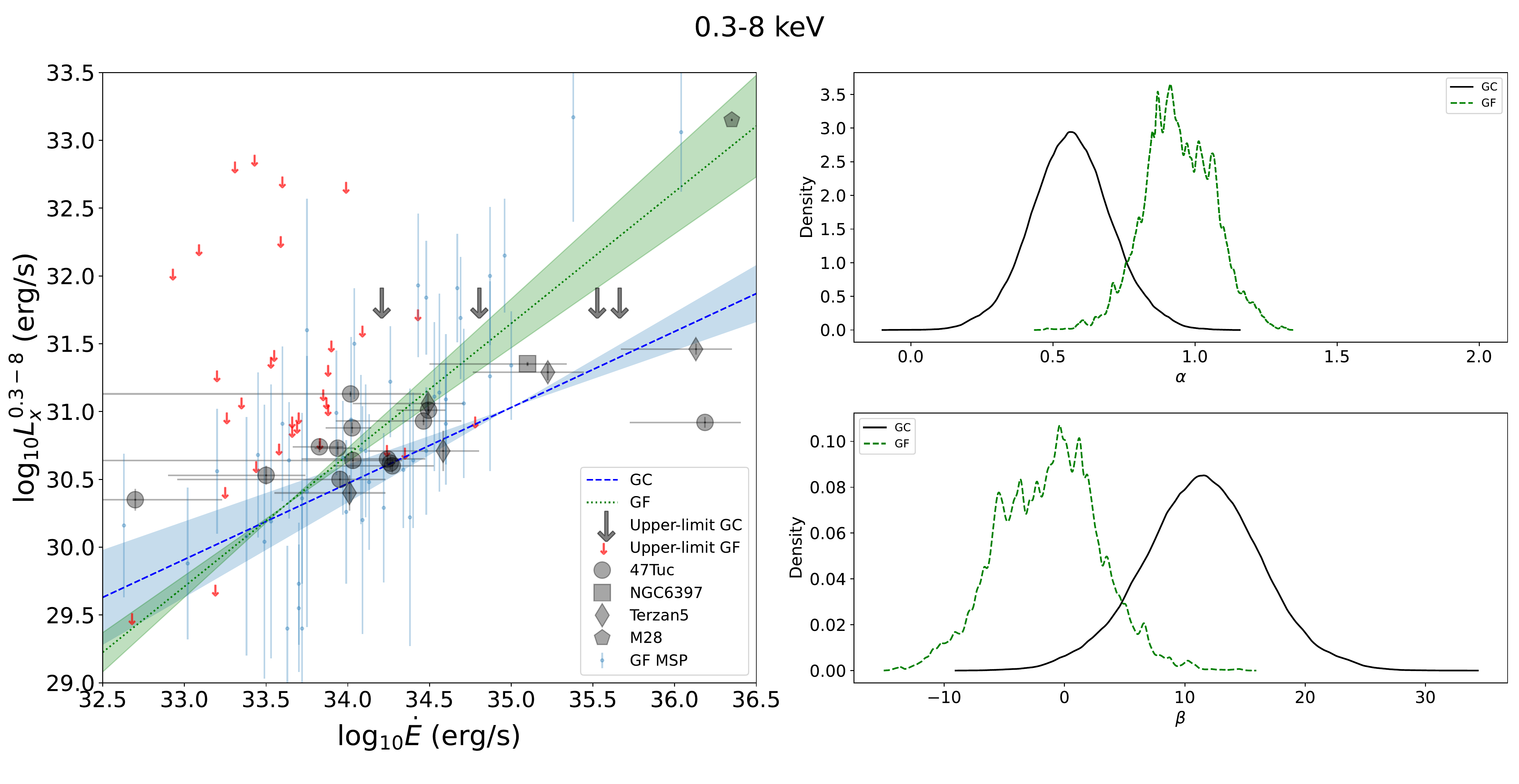}
\includegraphics[width=7in, angle=0]{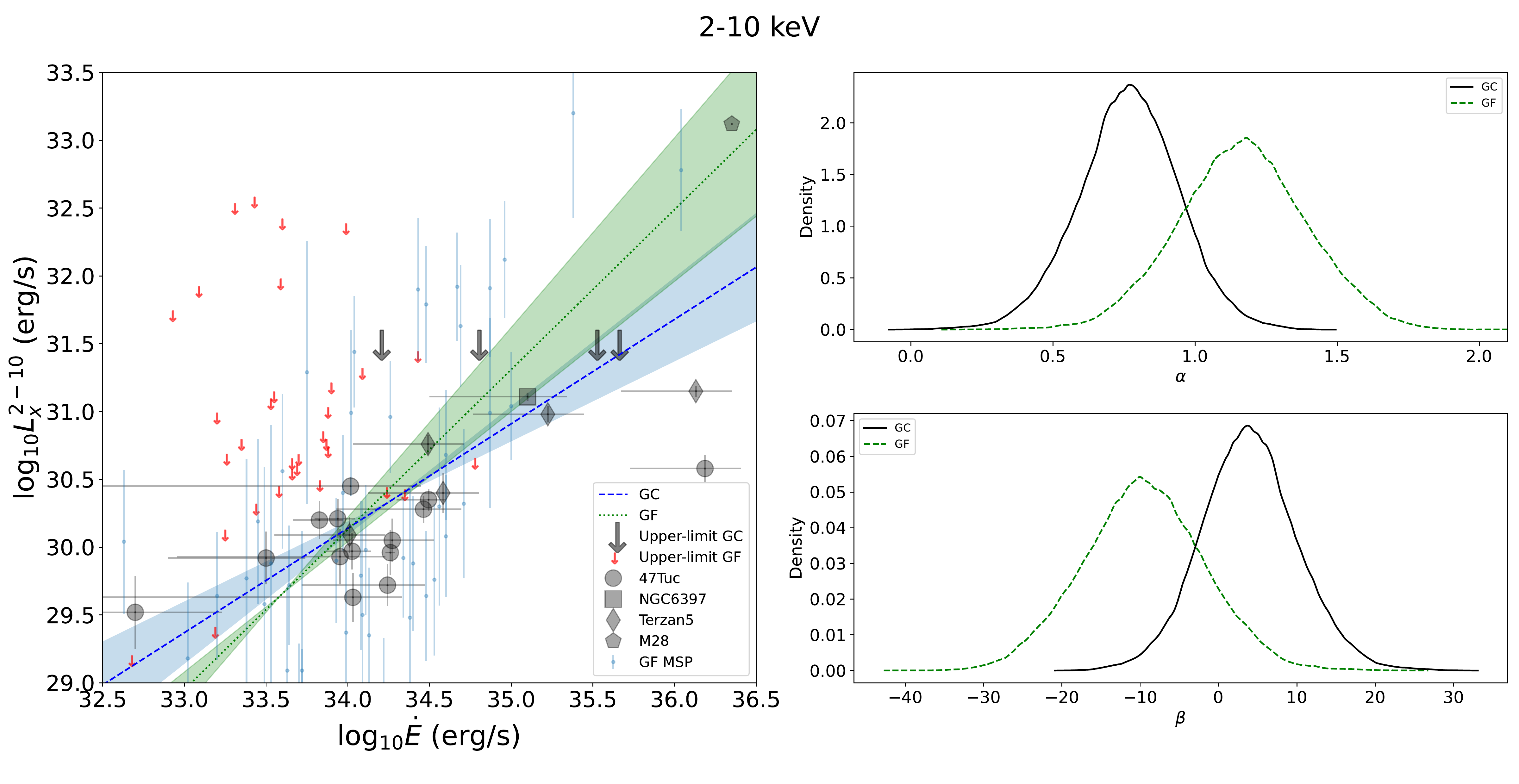}
\caption{\label{figure:figure4} {\it Upper-left} panel shows the plot of $L_{x}$ vs. $\dot{E}$ of GC MSPs in 0.3-8 keV for Group C. The blue dashed line and the corresponding shaded region show the best-fit linear model $\log L_{x}=\alpha\log\dot{E}+\beta$ as inferred from the censored data (i.e. confirmed detection + upper-limits) and the uncertainty respectively (cf. \autoref{tab:Table10}). The corresponding relation (i.e. green dotted line + shaded region) for the GF MSPs in the same band is plotted for comparison. Samples from different GCs are illustrated by different symbols. {\it Lower panel} shows the same plot but for the energy band 2-10 keV. Marginalized posterior probability distributions of the model parameters are shown in the {\it upper-right} and {\it lower-right} panels for 0.3-8~keV and 2-10~keV respectively.}
\end{figure}

\begin{figure}
\centering
\includegraphics[width=7in, angle=0]{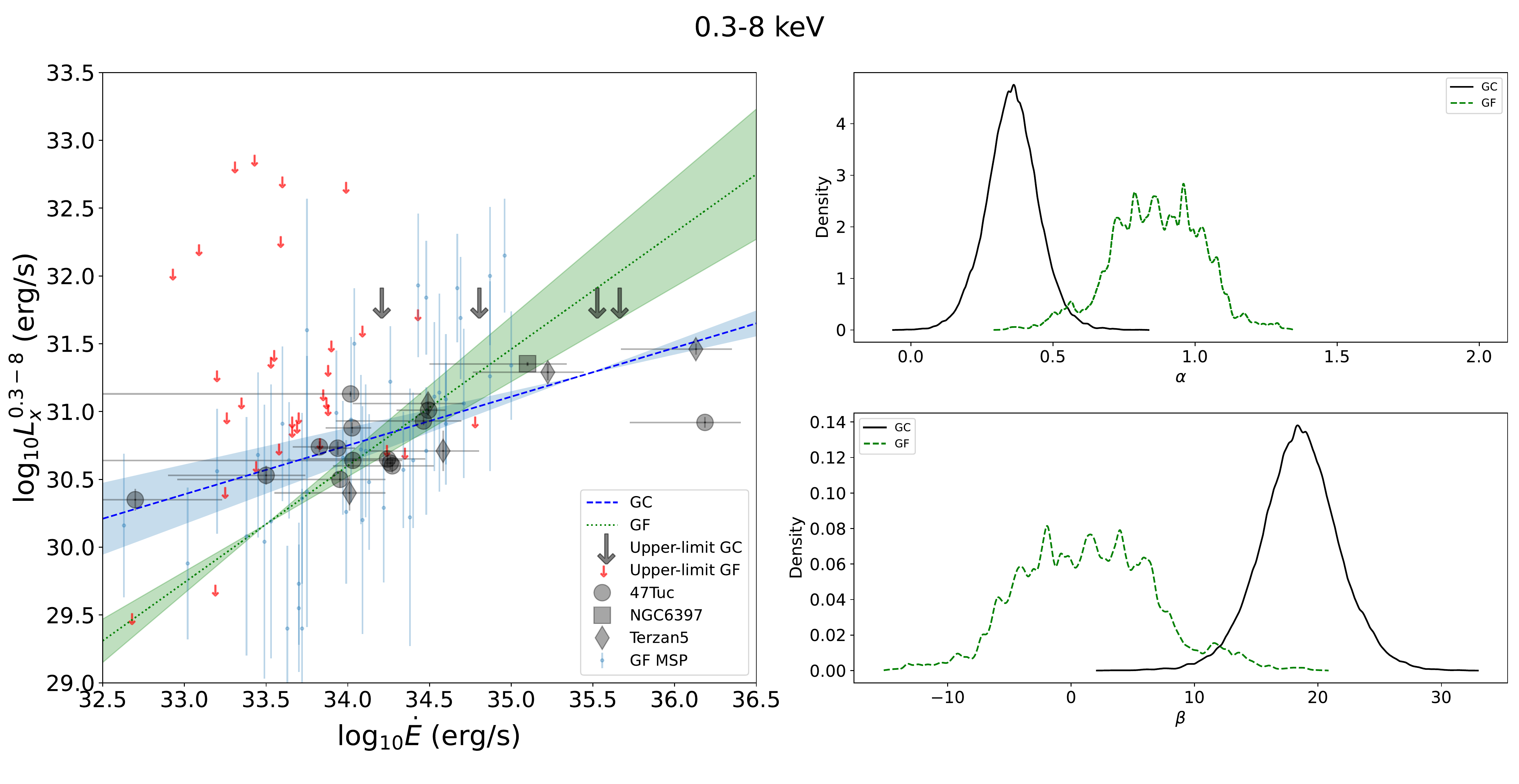}
\includegraphics[width=7in, angle=0]{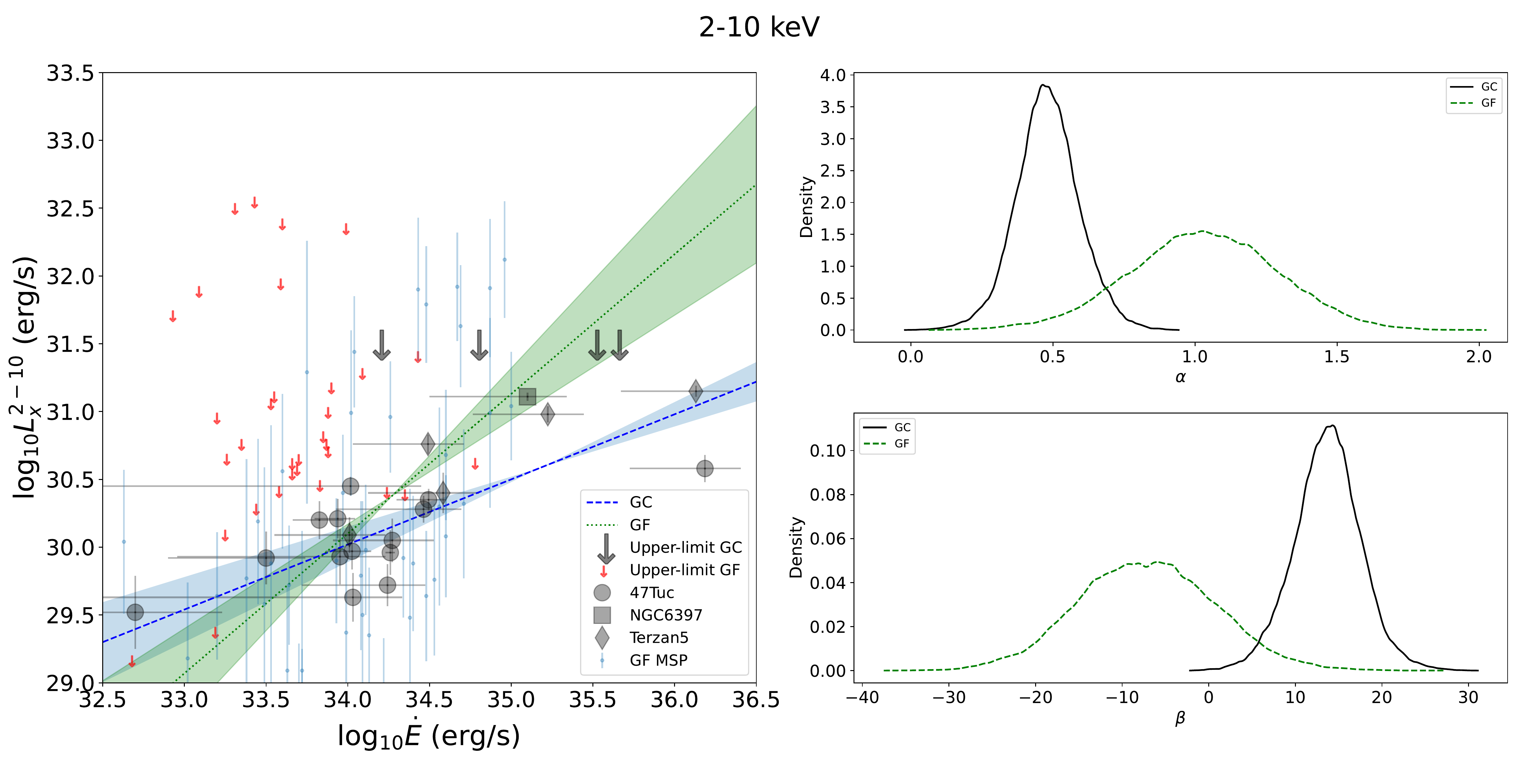}
\caption{\label{figure:figure5} Same as \autoref{figure:figure4} but with the outliers in both GCs (i.e. M28~A) and GF (i.e PSR~J0218+4232 and PSR~B1937+21) excluded.}
\end{figure}

\begin{figure}
\centering
\includegraphics[width=7in, angle=0]{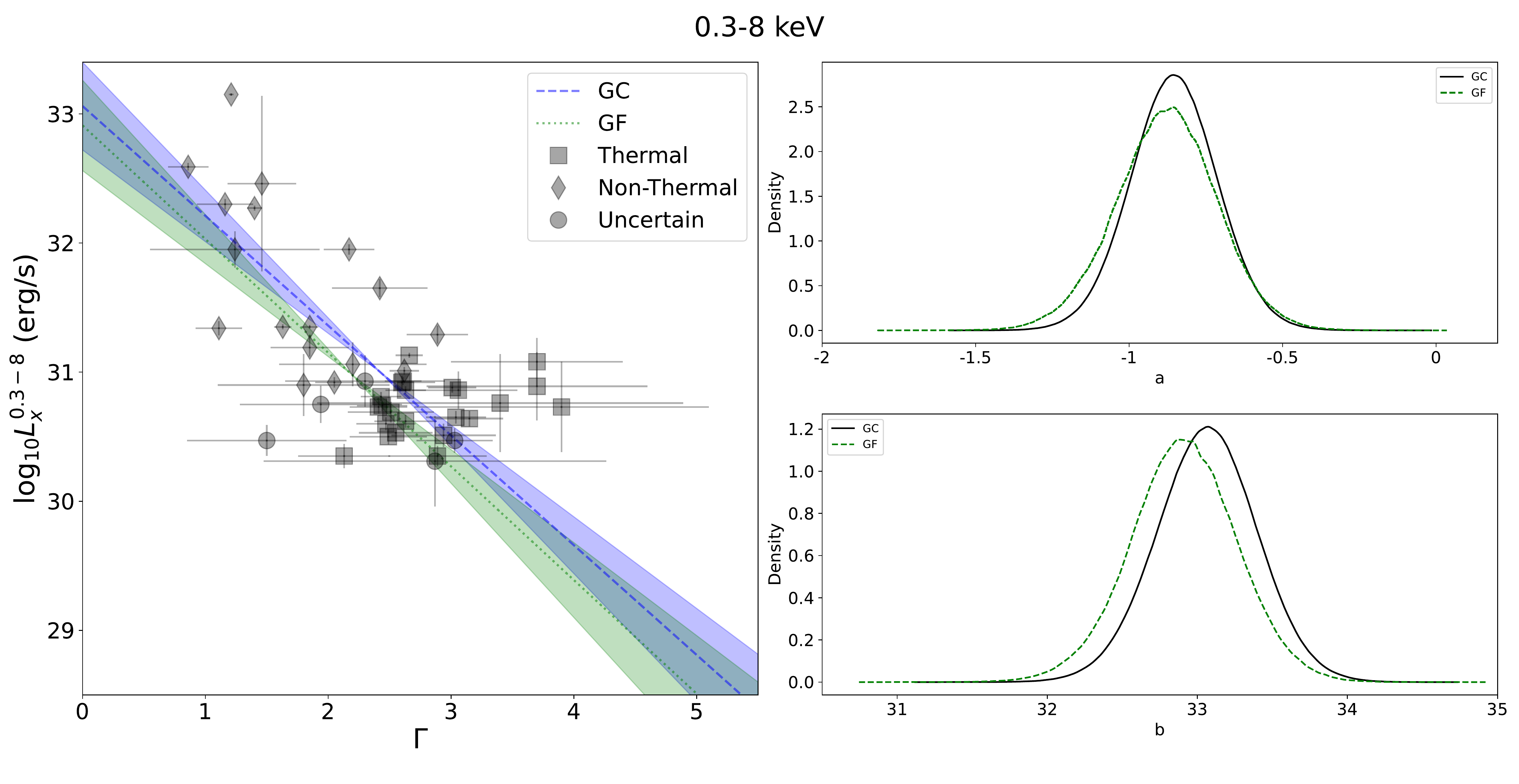}
\includegraphics[width=7in, angle=0]{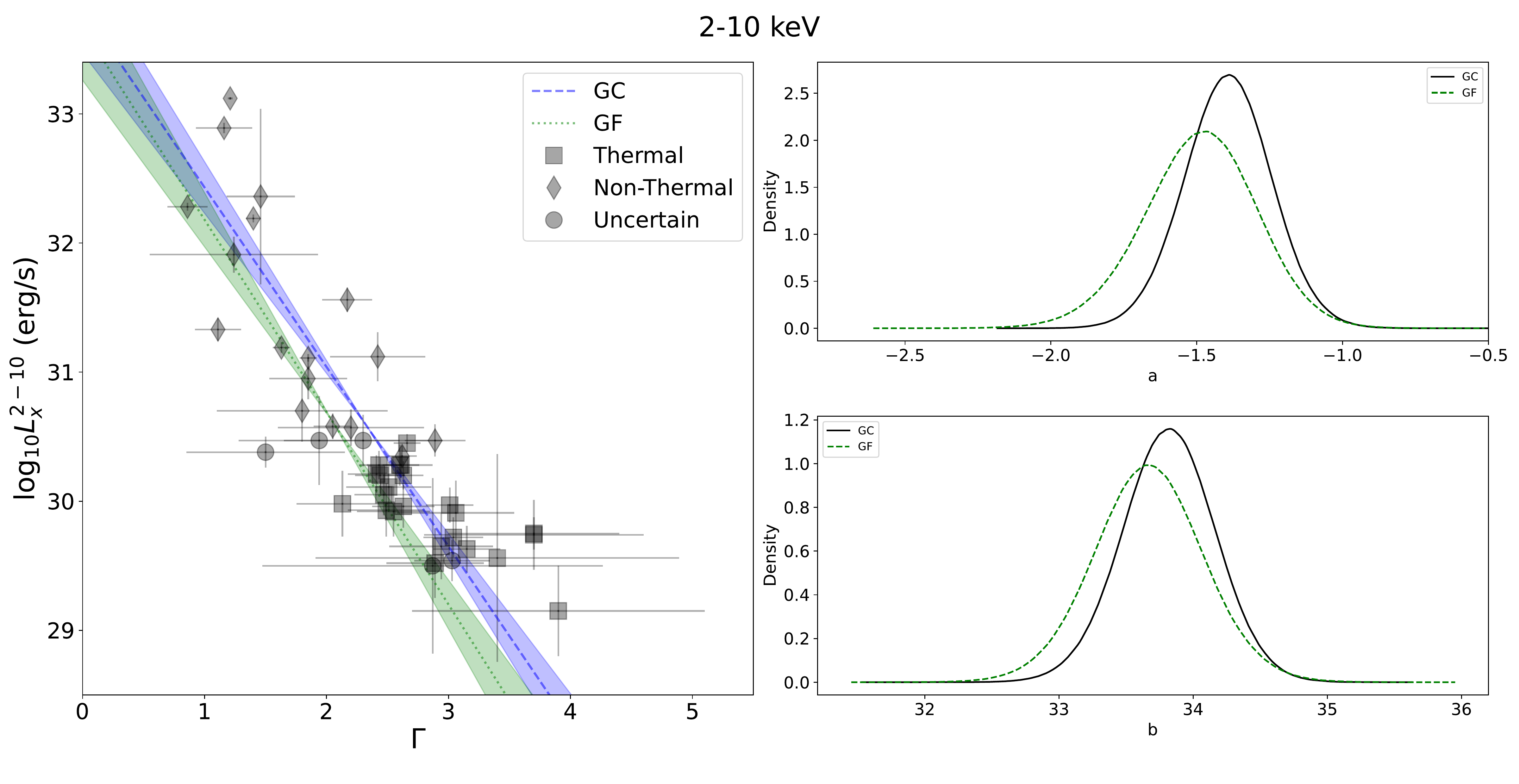}
\caption{\label{figure:figure6}  Correlation between $L_{x}$ and $\Gamma$ of X-ray detected GC MSPs in 0.3-8~keV ({\it Upper-left panel}) and 2-10~keV ({\it Lower-left panel}). 
The blue dashed line and the corresponding shaded region show the best-fit linear model $\log L_{x}=a\Gamma+b$ for GC MSPs as well as the uncertainty (cf. \autoref{tab:Table10}). 
The corresponding relation (i.e. green dotted line + shaded region) for the GF MSPs in the same band is plotted for comparison. 
Marginalized posterior probability distributions of the model parameters are shown in the {\it upper-right} and {\it lower-right} panels for 0.3-8~keV and 2-10~keV respectively.
}
\end{figure}

\begin{figure}
\centering
\includegraphics[width=7in, angle=0]{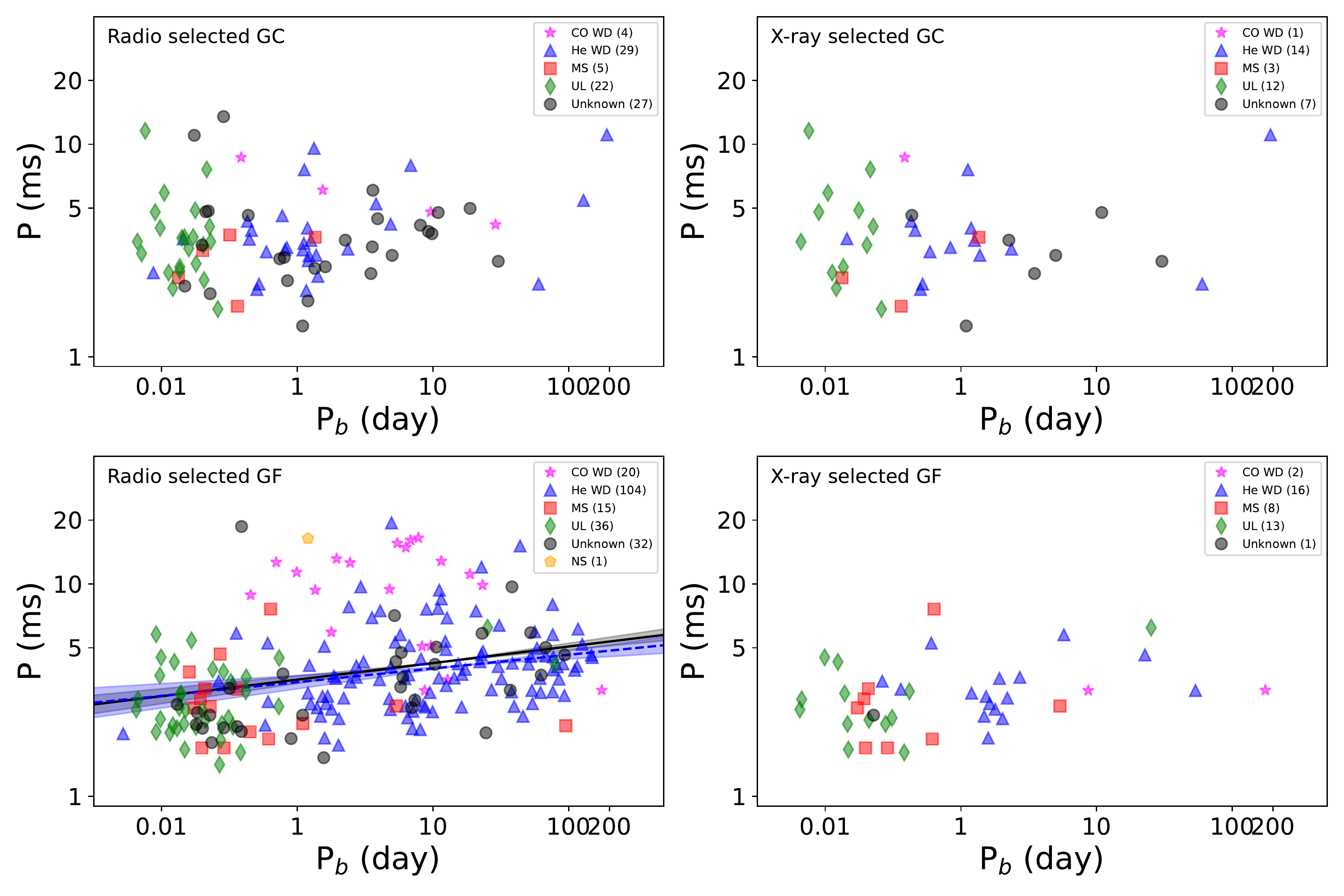}
\includegraphics[width=7in, angle=0]{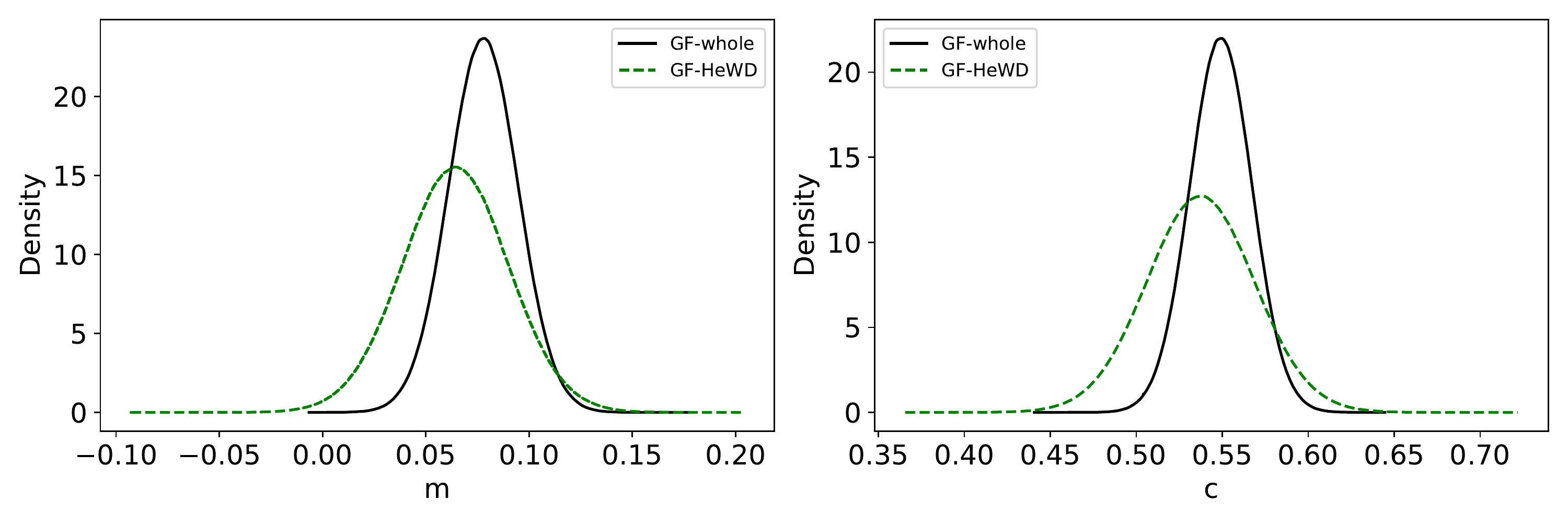}
\caption{\label{figure:figure7} The plots of $P_{b}$ vs. $P$ of radio/X-ray selected MSP binaries in GCs ({\it Top left/right panels}) and GF ({\it Middle left/right panels}). Systems with different companions are represented by different symbols. The bracketed numbers in the legends show the sample sizes in each class. The solid and the dashed lines in the {\it middle-left panel} represent the best-fit model $\log P=m\log P_{b}+c$ for the whole radio selected GF MSPs and those with He WD as companions respectively(cf. \autoref{tab:Table10}) . 
{\it Lower-left} and {\it lower-right panels}  show the marginalized posterior probability distributions for the model parameters.
}
\end{figure}

\begin{figure}
\centering
\includegraphics[width=7in, angle=0]{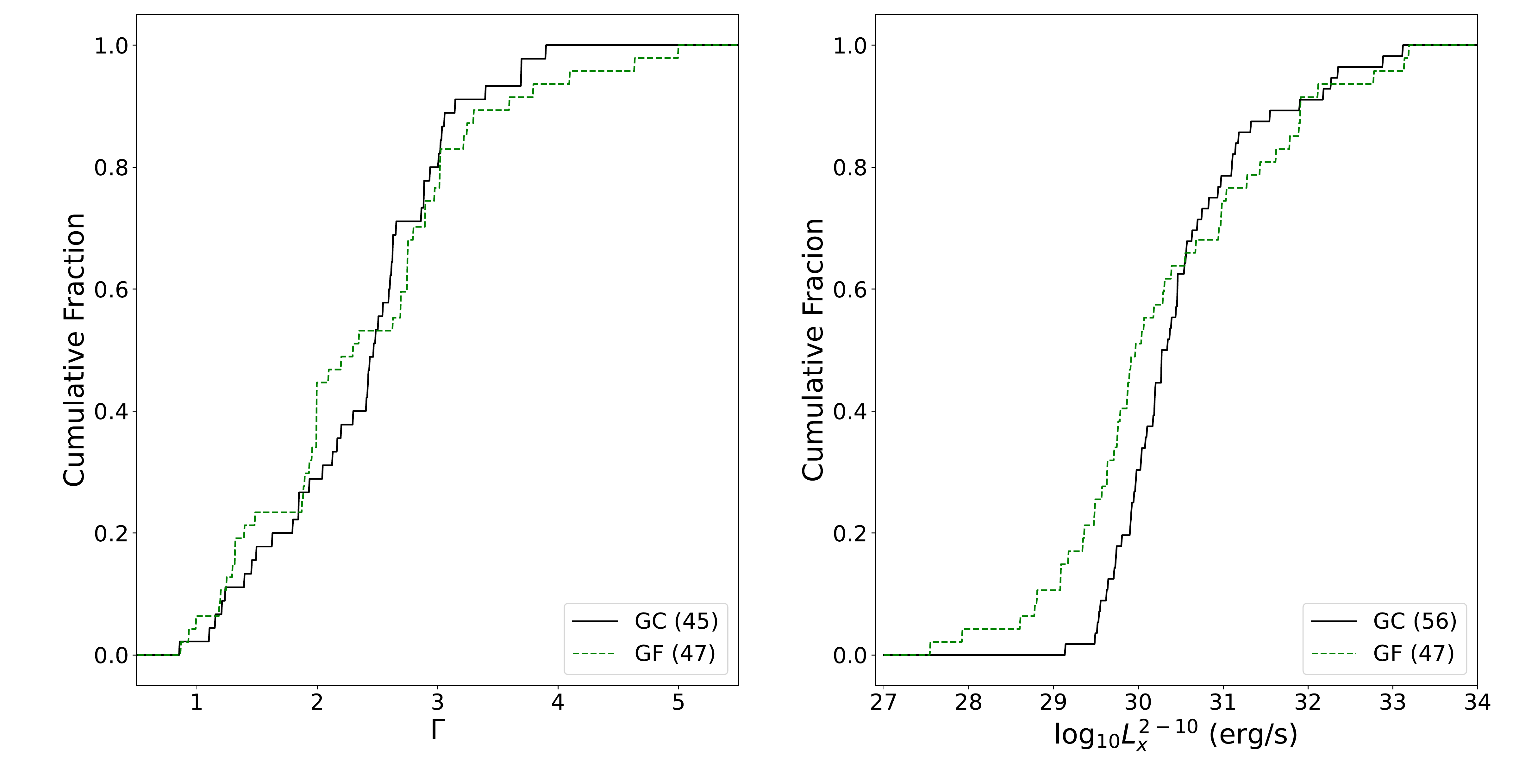}
\caption{\label{figure:figure8} Comparisons of eCDFs of $\Gamma$ and $L_{x}$ between GC MSPs and GF MSPs. The bracketed numbers in the legends show the sample sizes.}
\end{figure}

\begin{figure}
\centering
\includegraphics[width=7in, angle=0]{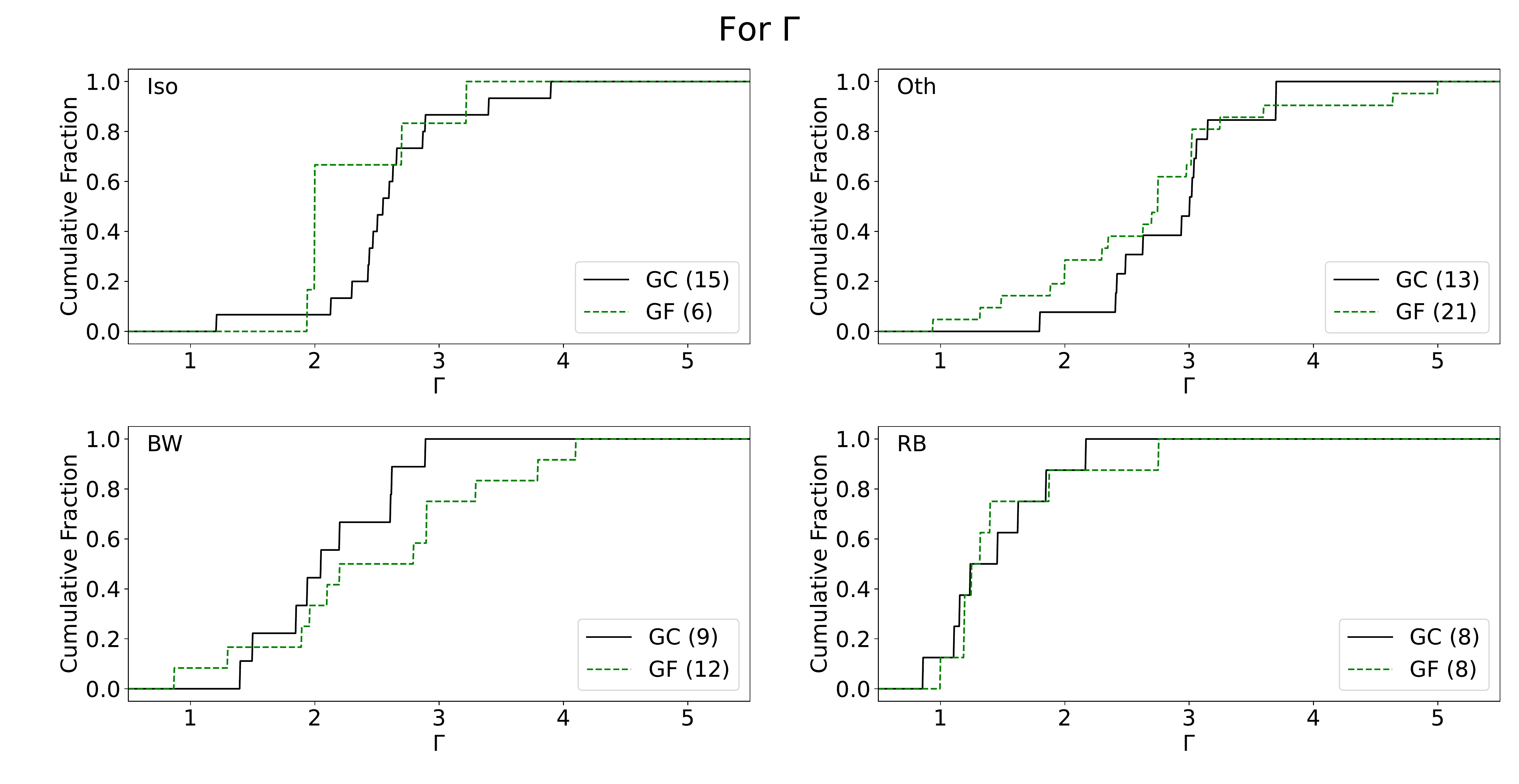}
\includegraphics[width=7in, angle=0]{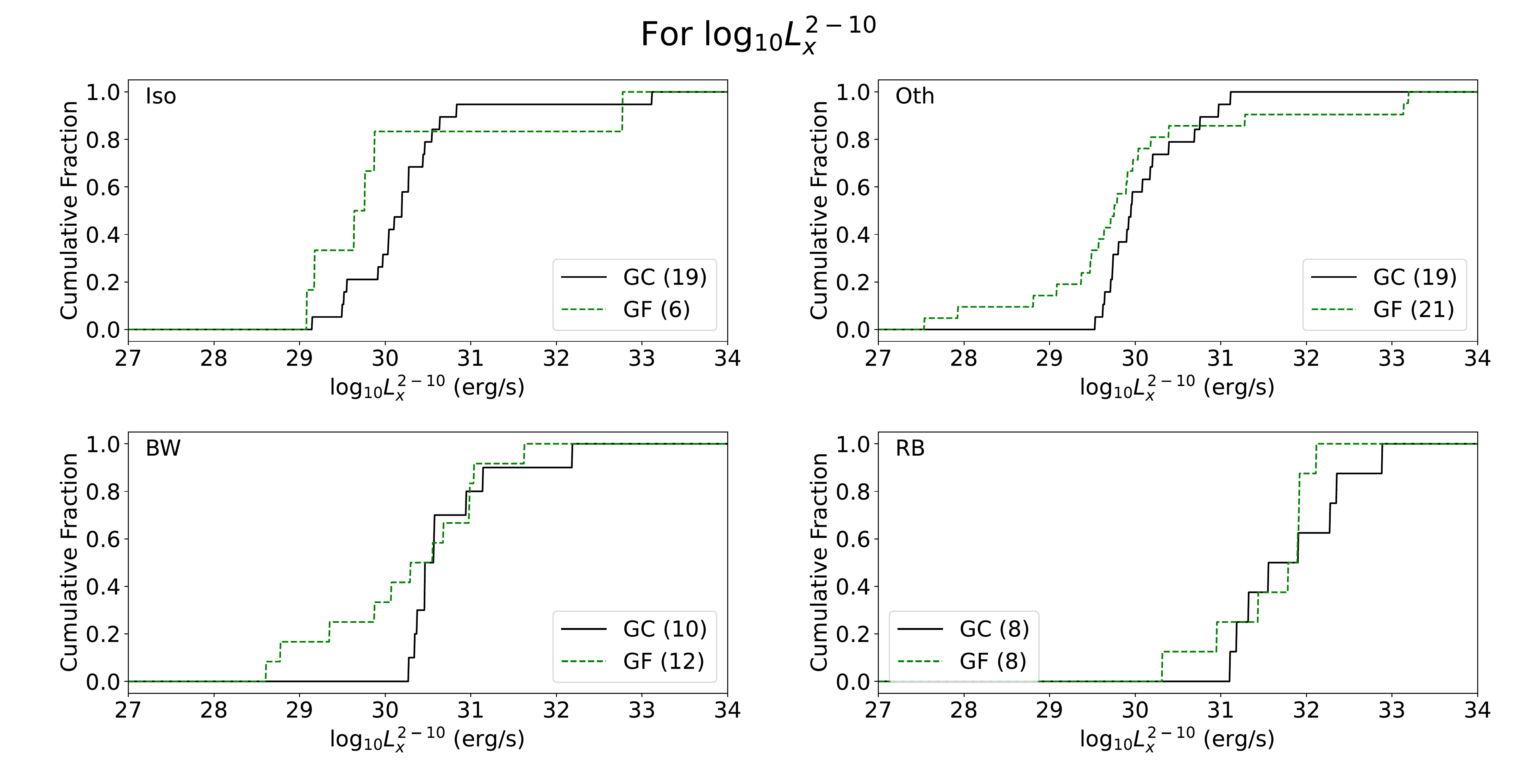}
\caption{\label{figure:figure9} Comparisons of eCDFs of $\Gamma$ ({\it Upper 2$\times$2 panels}) and $L_{x}$ ({\it Lower 2$\times$2 panels}) for the corresponding classes between GC MSPs and GF MSPs.}
\end{figure}

\begin{figure}
\centering
\includegraphics[width=7in, angle=0]{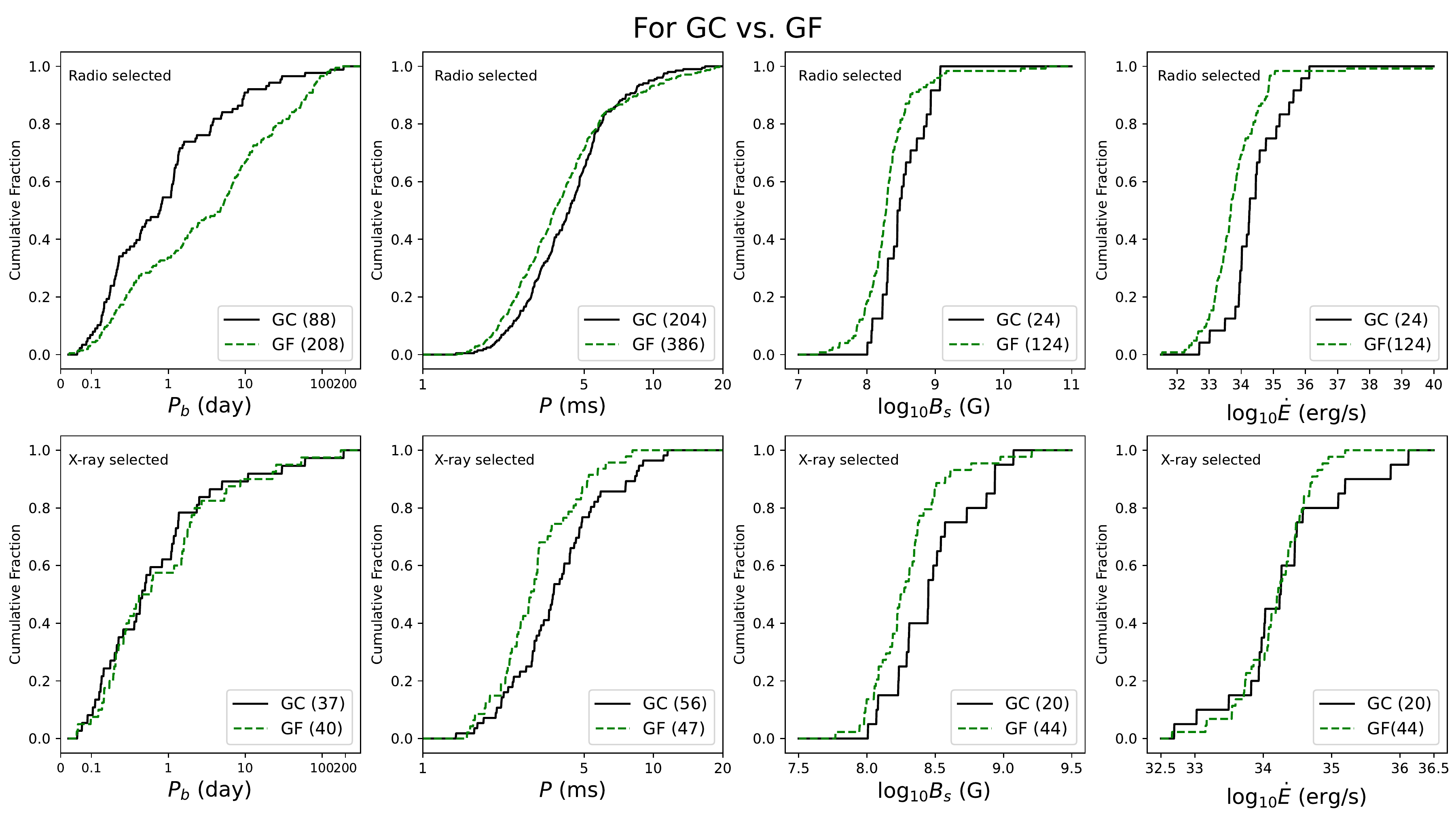}
\includegraphics[width=7in, angle=0]{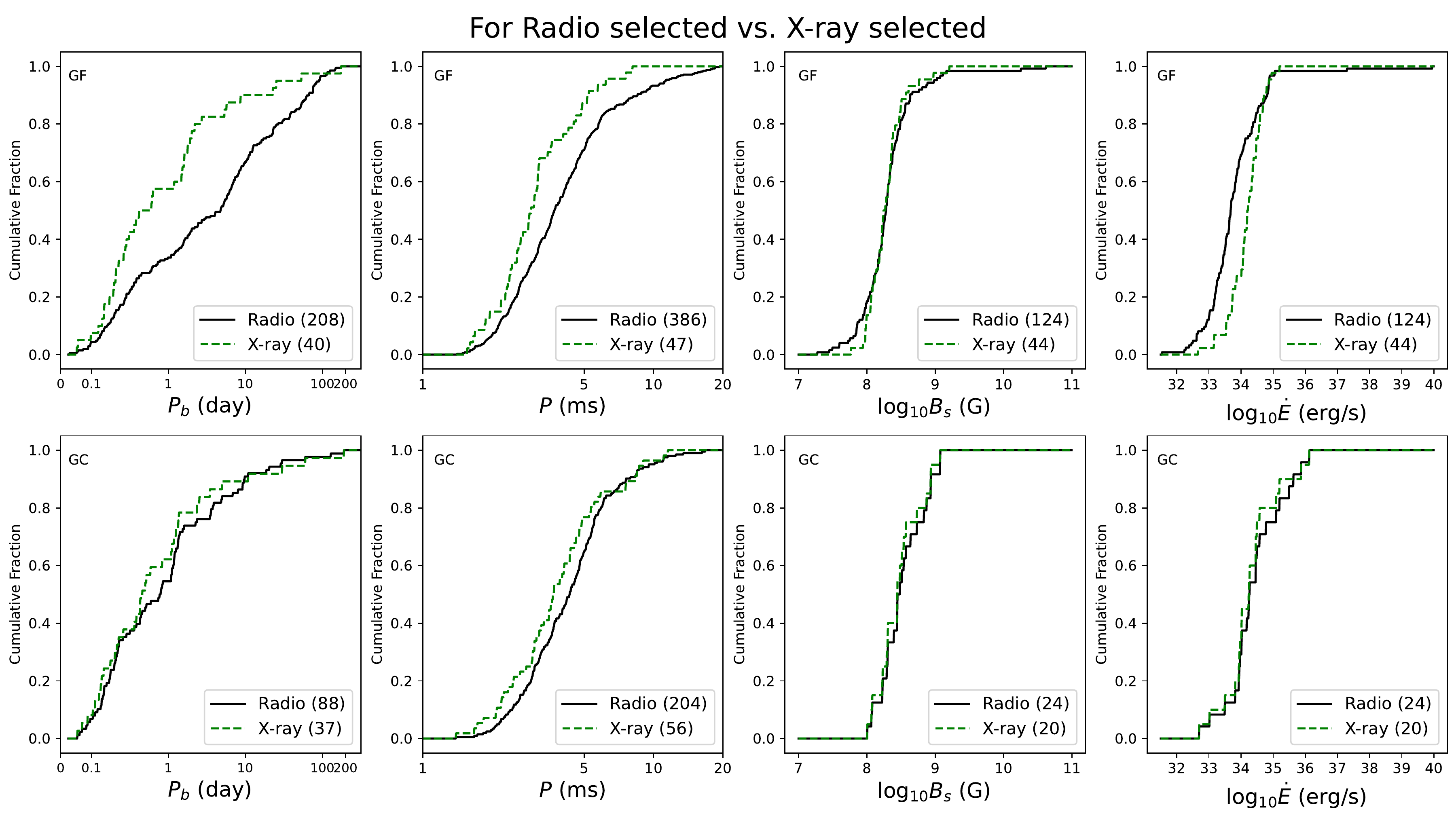}
\caption{\label{figure:figure10} ({\it First row}) Comparisons of eCDFs of $P_{b}$, $P$, $B_{s}$ and $\dot{E}$ between radio selected MSPs in GCs and GF. The outliers in both GCs (i.e. M28~A) and GF (i.e. PSR~J0218+4232 and PSR~B1937+21) are excluded in comparing $B_{s}$ and $\dot{E}$. ({\it Second row}) Comparisons of the same set of parameters between X-ray selected MSPs in GCs and GF. ({\it Third row}) Comparisons the same set of parameters between X-ray selected and radio selected MSPs in GF. ({\it Fourth row}) Comparisons the same set of parameters between X-ray selected and radio selected MSPs in GCs.}
\end{figure}

\begin{figure}
\centering
\includegraphics[width=7in, angle=0]{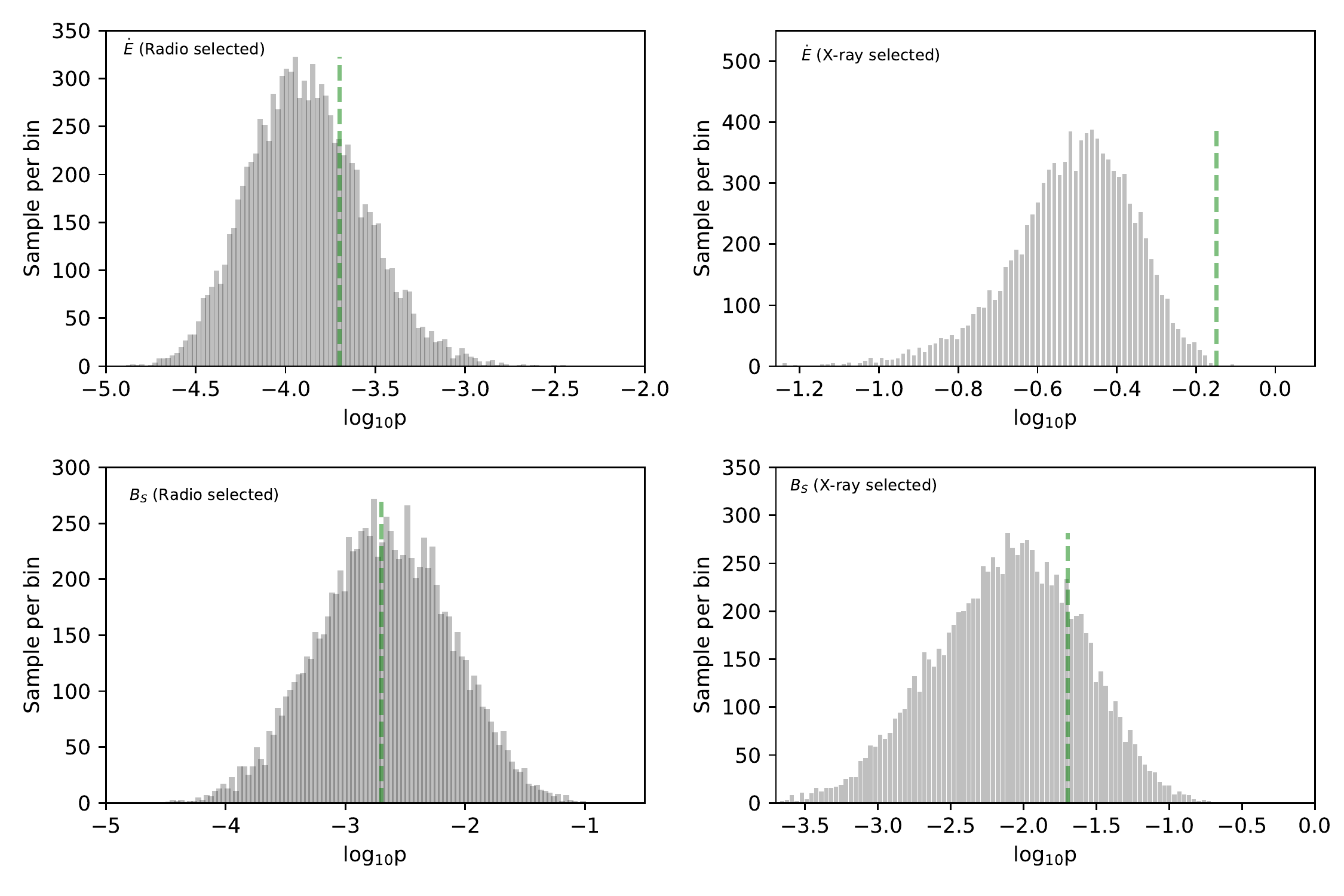}
\caption{\label{figure:figure11} Empirical distributions of null hypothesis probabilities (i.e. $p-$values) resulted from the A-D test in comparing 10000 sets of simulated samples of $\dot{E}$ and $B_{s}$ drawn from both GC and GF samples with outliers excluded. The green dashed lines illustrate the $p-$values obtained from comparing the observed samples between GC and GF.}
\end{figure}

\begin{figure}
\centering
\includegraphics[width=7in, angle=0]{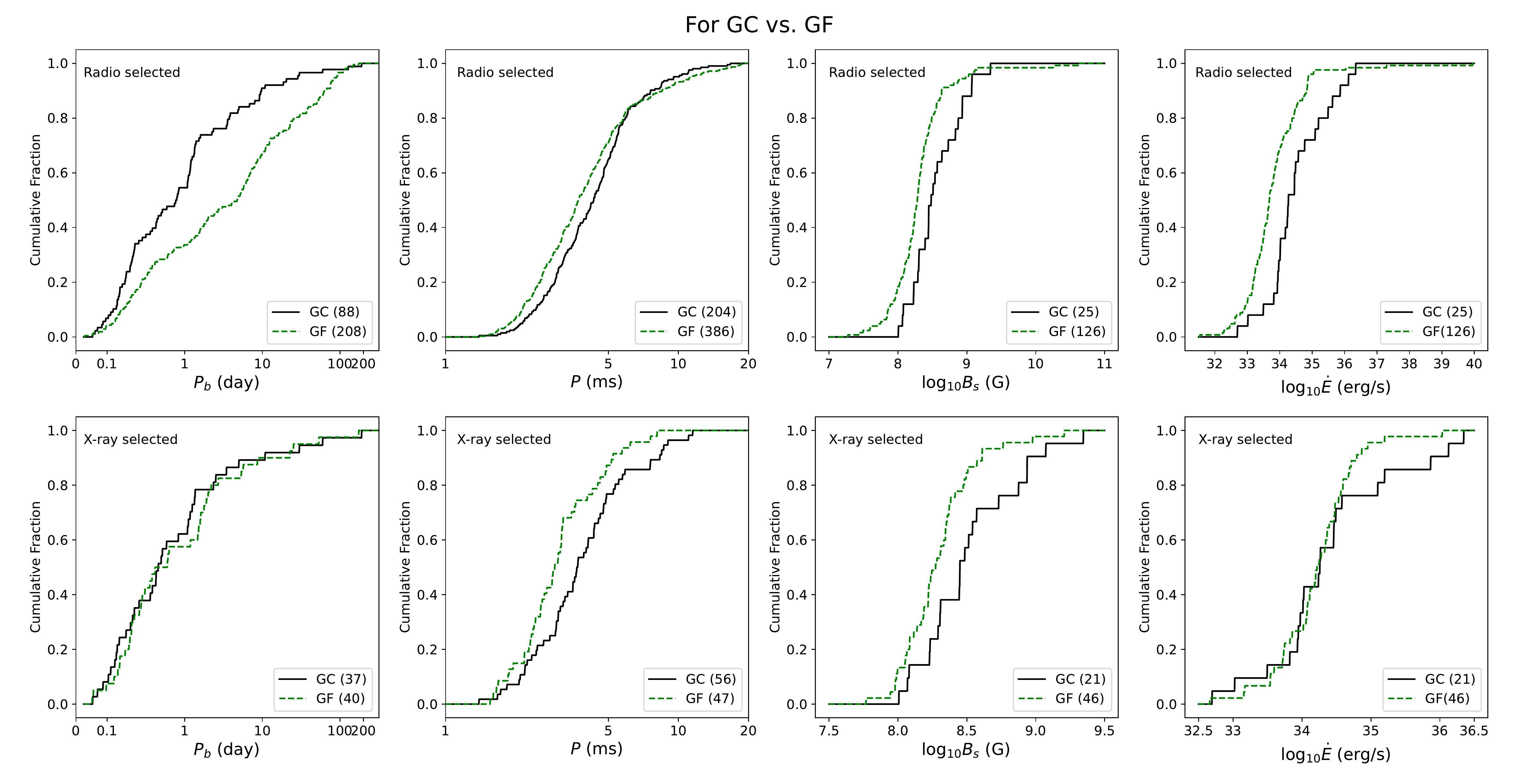}
\includegraphics[width=7in, angle=0]{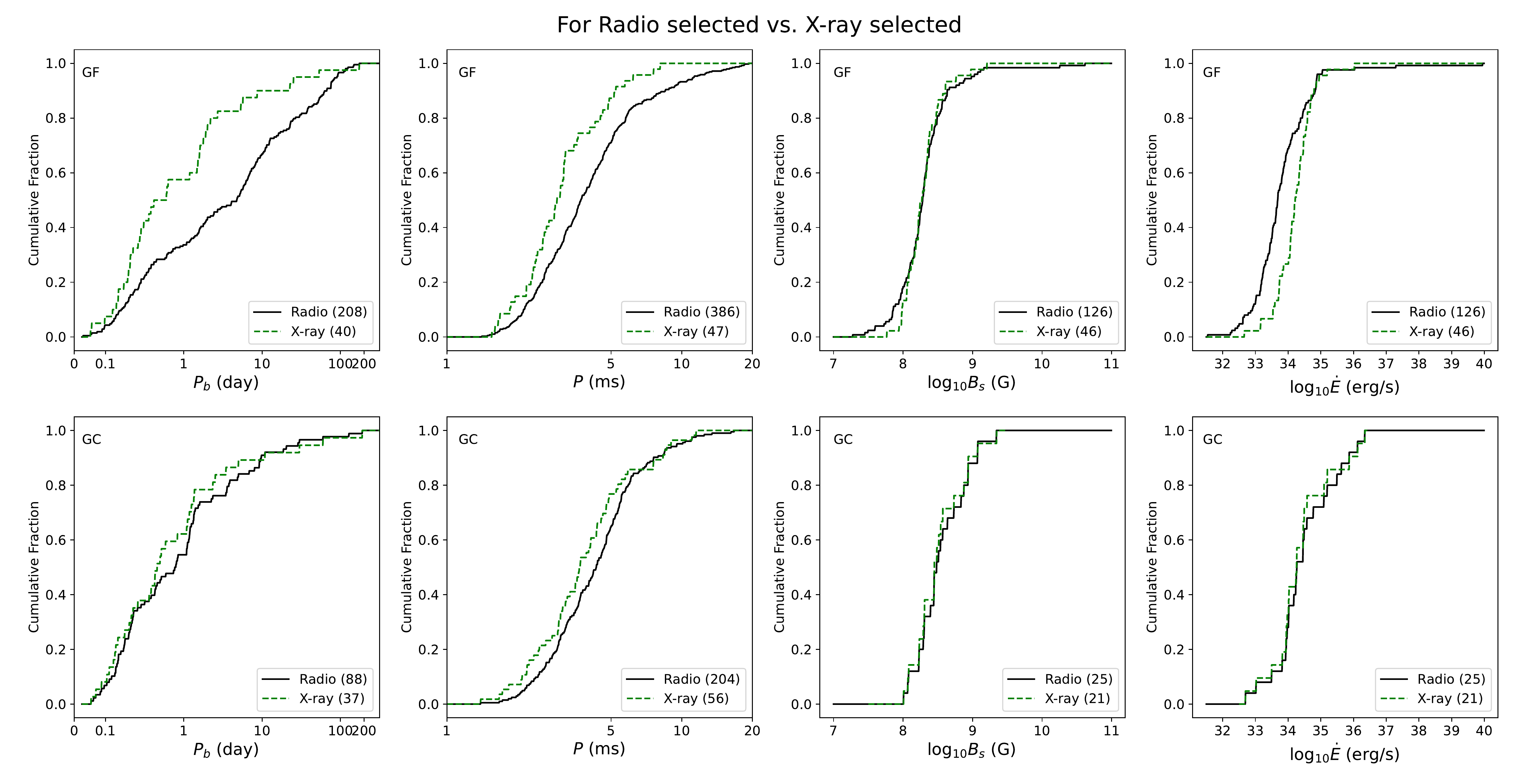}
\caption{\label{figure:figure12} Same as \autoref{figure:figure10} but with M28~A, PSR~J0218+4232 and PSR~B1937+21 included for comparing $B_{s}$ and $\dot{E}$.}
\end{figure}

\begin{figure}
\centering
\includegraphics[width=7in, angle=0]{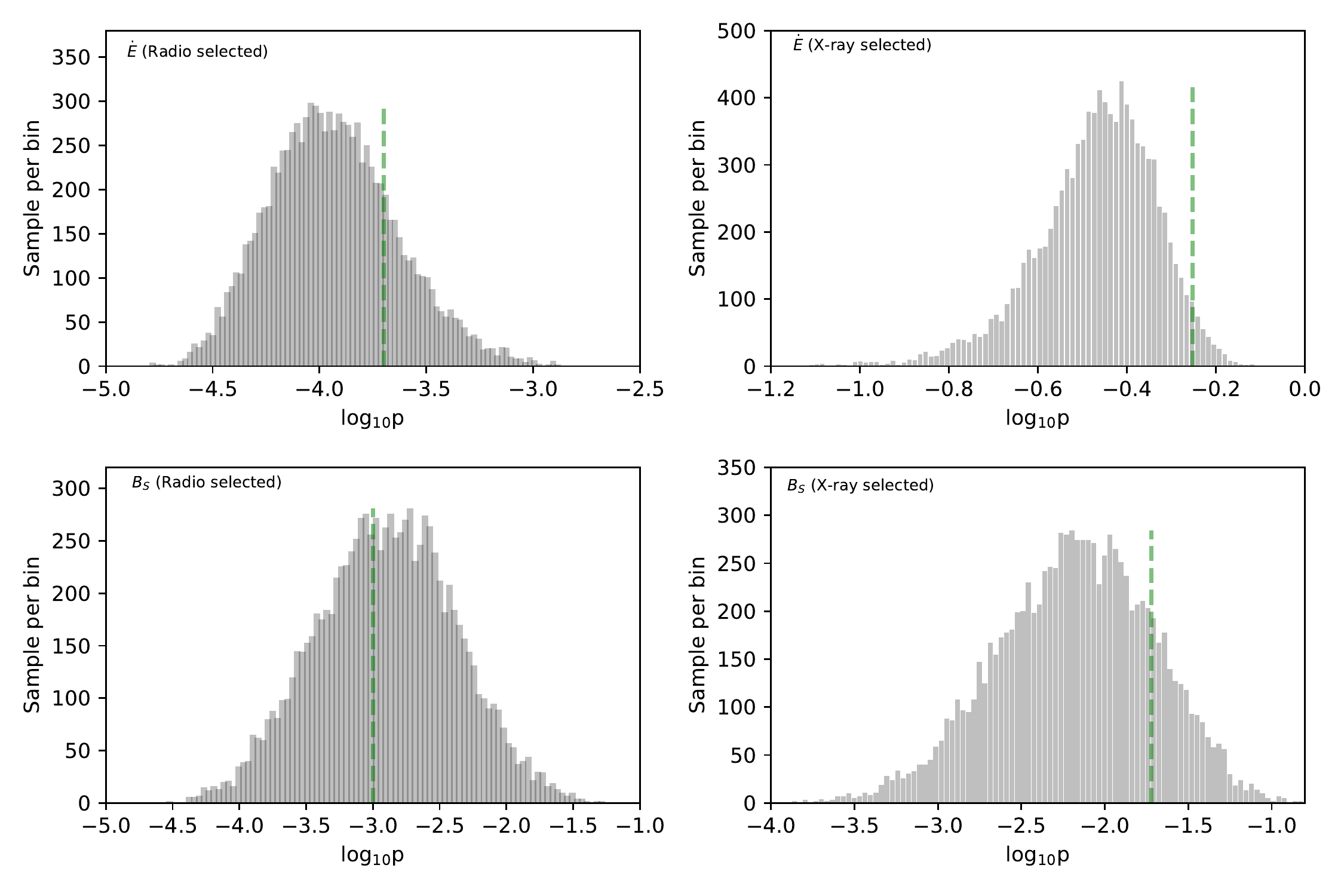}
\caption{\label{figure:figure13} Same as \autoref{figure:figure11} but with M28~A, PSR~J0218+4232 and PSR~B1937+21 included.}
\end{figure}

We also notice that the fraction of isolated MSPs in GCs is larger than that in the GF, which is particularly obvious in the X-ray selected sample (\autoref{figure:figure1}). \cite{Verbunt2014.561} have suggested that the large fraction of isolated MSPs in GCs can be a result of dynamical disruption. Although this is physically plausible, we would like to point out that this might also be an observational effect. In the GF, the X-ray counterparts of MSPs are detected by pointed observations towards individually chosen targets. This can lead to a selection bias towards those bright sources with interesting behavior such as spider pulsars. This can possibly account for their large fraction in the GF population and hence suppress the proportion of isolated MSPs. On the other hand, there is no such bias in searching X-ray counterparts of MSPs in GCs since all MSPs in a given GC are observed at once in the X-ray image. In view of this, we cannot exclude the possibility that the larger fraction of isolated MSPs in GCs is a result of observational bias. For resolving this issue, a systematic all-sky X-ray imaging survey on the GF will be needed (e.g. with eROSITA). With a less biased sample, the proportions of different classes of X-ray emitting MSPs can be re-examined.

For the radio selected samples, the larger fraction of isolated MSPs in GCs can also be a result of observational bias. Detecting MSPs in binaries is more challenging than detecting isolated MSPs because searches of orbital parameters are also required. For GCs, the situation is exacerbated by the intracluster acceleration. Any deviation of the timing solutions from the actual values might lead to smearing of the radio pulses which can hamper the detection. As a result, this can possibly lead to a larger proportion of isolated MSPs in GCs. Therefore, the conclusion of whether the dynamical disruption in GCs can lead to more isolated MSP is not without ambiguity.

During recycling, accretion on the neutron stars can induce the decay of the surface magnetic field through the processes such as Hall effect and Ohmic dissipation \citep{Cumming2004ApJ.609.999C}. Therefore, perturbation on the spin-up process by the dynamical interactions can halt the magnetic field decay in GC MSPs. This is consistent with our findings that $B_{s}$ of GC MSPs are larger than those in the GF (See \autoref{tab:Table8} \& \ref{tab:Table9}). This inference has also been reported by \cite{Verbunt2014.561} and \cite{Konar2010.409}. However, the ways how these studies collected their samples are different from our approach. It is unclear whether the estimates of $\dot{P}$ adopted in these previous works have the acceleration terms corrected. It appears that these studies have collected those have larger $\dot{P}_{\rm obs}$ so as to have a smaller fractional contamination attributed by the cluster acceleration. However, this unavoidably introduced the bias that favors the conclusion that $B_{s}$ of GC MSPs is higher. 

In our study, a majority of our samples of $B_{s}$ are estimated by the $\dot{P}$ with their acceleration terms corrected by long-term pulsar timing (see \autoref{tab:Table4} and Section 4). Such correction is unlikely suffered from the aforementioned bias. However, as the measurement of time derivative of the orbital period $\dot{P}_{b}$ should be easier for large $P_{b}$, most of the systems with intrinsic $\dot{P}$ estimated by this method are non-spider MSP binaries. This can introduce another bias in this comparison as we do not know the $B_{s}$ for the spider and isolated MSPs. On the other hand, \cite{Lee2018.864} found that all different types of MSPs in the GF have similar $B_{s}$ \citep[see Figure 6 in][]{Lee2018.864}. If this were also the case in GCs, our inference might remain be valid.   

To understand the cause of the selection effect imposed by X-ray observations, we need to discuss the spin-down power $\dot{E}$ of MSPs. In the GF, X-ray observations apparently pick the more powerful MSPs (i.e. larger $\dot{E}$) which are more luminous in X-ray as $L_{x}\propto\dot{E}^{1.31}$ \citep{Lee2018.864}. Since the distributions of $B_{s}$ of both radio/X-ray selected samples in GF are similar and $\dot{E}$ is proportional to $B_{s}^{2}/P^{4}$, this explains why the X-ray emitting MSPs in GF generally rotate faster. Furthermore, because of the correlation between $P$ and $P_{b}$, this also naturally explains why the X-ray emitting MSP binaries have tighter orbits in the GF. 

However, in contrast to the situation in the GF, we do not find any evidence for the X-ray selection effect on the GC MSPs (see \autoref{tab:Table8} \& \ref{tab:Table9}). We speculate that this might be accounted by the fact that the GC MSPs follow a different $L_{x}-\dot{E}$ relation. In our adopted sample, we found $L_{x}\propto\dot{E}^{0.4-0.8}$ (\autoref{tab:Table10}). It appears that the $L_{x}$ of the GC MSPs have a less sensitive dependence on $\dot{E}$ than those in the GF. This might explain why the selection effect on the GC population is less prominent than that in the GF. 

For the GC MSPs, it is interesting to notice that the inferred dependence of their $L_{x}$ on $\dot{E}$ is consistent with that of Goldreich-Julian current $J_{GJ}\propto\sqrt{\dot{E}}$ \citep{Goldreich.Julian1969ApJ.157.869G}. This suggests that the X-rays are likely resulted from the polar cap heating by the back-flow current, which should be scaled with $J_{GJ}$. Since there is an indication that $B_{s}$ is stronger in GC MSPs, this might facilitate the magnetic pair creation close to the stellar surface and result in a higher efficiency of polar cap heating than their GF counterparts \citep{Cheng2003.598}. Also, \cite{Takata2010ApJ.715.1318T} suggest that if these magnetic pairs stream back to the outermagnetosphere, they will restrict the size of the outergap accelerator. And hence the production of non-thermal emission will be limited. This is consistent with the fact that the X-rays from almost all the GC MSPs adopted in examining the  $L_{x}-\dot{E}$ relation are thermal dominant (\autoref{tab:Table2}). 

Lastly, we would like to highlight that the $L_{x}-\dot{E}$ relation for the GC MSPs reported in this work can be biased by the way we collected the sample. Since most of sample in \autoref{tab:Table4} have the acceleration terms in their observed $\dot{P}$ corrected by $\dot{P}_{b}$, it is biased by those having large $P_{b}$ which are mostly non-spider MSP binaries. While we found that $\dot{E}$ of the X-ray selected MSPs are comparable in GCs and GF, we are not sure how does this conclusion will be altered if the isolated and spider MSPs in GCs are included. In the GF, \cite{Lee2018.864} have shown that the $\dot{E}$ of isolated and spider MSPs are comparable with each other and they are much larger than the non-spider MSPs \citep[see Figure 7 and Table 4 in ][]{Lee2018.864}. If the situation is similar in GC population, this might suggest that GC MSPs are more powerful than those in the GF. While measuring the line-of-sight acceleration of isolated and spider MSPs by long-term radio timing can be challenging, a systematic analysis of the whole GC MSP population with the mean-field acceleration and numerical simulation are encouraged for further investigation.

Recently, a new sub-class of BWs in the GF which are referred as {\it Tidarren} systems has been identified \citep{Romani2016ApJ.833.138R}. While classic BWs have companion mass in a range of $\sim0.02-0.05M_{\odot}$, the companions of Tidarren systems have mass $<0.015M_{\odot}$ and with their hydrogen completely stripped off by the powerful pulsar wind. It has been suggested that they are descendants of the ultracompact X-ray binaries and follow a different evolutionary path. In a recent kinematic analysis of two Tidarren systems in the GF, \cite{Long2022ApJ.934.17} have shown that such systems can be originated from GCs. With more Tidarren systems discovered in the future, one can further examine the possible intricate relation between the MSPs in GCs and the GF.

During the reviewing process, we became aware that a publication by \cite{Zhao2022MNRAS.511.5964Z} on a similar subject as our work. The authors have also presented a population analysis of X-ray properties of GC MSPs but with a focus different from our work. While we have performed a systematic comparison between the MSPs in GCs and GF to explore the influence of dynamical interactions and the selection effects imposed by X-ray observations, \cite{Zhao2022MNRAS.511.5964Z} have focused solely on the X-ray properties of GC MSPs (e.g. examining their X-ray luminosity functions, placing upper and lower bounds on the number of MSPs in various GCs). 
On the other hand, their independent work allows us to cross-check $L_{x}$ and we found that the estimates in both works are consistent.

\acknowledgments
The authors would like to thank Jongsuk Hong for the valuable comments and suggestions.
J.L. is supported by the National Research Foundation of Korea grant 2016R1A5A1013277, 
 2022R1F1A1073952 and
National Research Foundation of Korea grant funded by the Korean Government 
(NRF-2019H1A2A1077350-Global Ph.D. Fellowship Program);
C.Y.H. is supported by the research fund of Chungnam National University and by the National Research Foundation of Korea grant 2022R1F1A1073952.
J.T. are supported by the National Key Research and Development Program of China (grant No. 2020YFC2201400) and the National Natural Science Foundation of China (NSFC, grant No. 12173014). 
A.K.H.K. is supported by the National Science and Technology Council of Taiwan through grant 111-2112-M-007-020.
P.H.T. is supported by the NSFC grant No. 12273122 and the China Manned Space Project (No. CMS-CSST-2021-B09).
K.L.L. is supported by the National Science and Technology Council of the Republic of China (Taiwan) through grant 111-2636-M-006-024, and he is also a Yushan Young Fellow supported by the Ministry of Education of the Republic of China (Taiwan).

\bibliography{reference}
\bibliographystyle{aasjournal}

\clearpage

\begin{table}[h]
    \caption{\label{tab:Table1} List of {\it Chandra} ACIS observations of GCs with MSPs identified.}
    \scriptsize
    \begin{center}
    \begin{tabular}{c c c c}
    \hline
    \hline
    Obs.ID & Start Date and Time & Inst. & Exp. time \\
    & (UTC) & (ACIS-) & (ks) \\
    \hline
    \multicolumn{4}{c}{NGC1851} \\
    \hline
    8966 & 2008-04-04T15:32:24 & S & 18.80 \\
    15734 & 2015-02-04T21:01:44 & S & 19.80 \\
    17588 & 2015-02-07T13:28:34 & S & 27.67 \\
    \hline
    \multicolumn{4}{c}{NGC5139 ($\omega$ Cen)} \\
    \hline
    653 & 2000-01-24T02:13:58 & I & 25.03 \\
    1519 & 2000-01-25T04:32:36 & I & 43.59 \\
    13727 & 2012-04-16T06:18:36 & I & 48.53 \\
    13726 & 2012-04-17T08:16:43 & I & 173.74 \\
    \hline
    \multicolumn{4}{c}{NGC5272 (M3)} \\
    \hline
    4543 & 2004-05-09T17:26:32 & S & 10.15 \\
    \hline
    \multicolumn{4}{c}{NGC5904 (M5)} \\
    \hline
    2676 & 2002-09-24T06:51:22 & S & 44.66 \\
    \hline
    \multicolumn{4}{c}{NGC6121 (M4)} \\
    \hline
    946 & 2000-06-30T04:24:23 & S & 25.82 \\
    7447 & 2007-07-06T05:26:35 & S & 45.46 \\
    7446 & 2007-09-18T02:47:24 & S & 47.93 \\
    \hline
    \multicolumn{4}{c}{NGC6205 (M13)} \\
    \hline
    7290 & 2006-03-09T23:01:13 & S & 27.89 \\
    5436 & 2006-03-11T06:19:34 & S & 26.80 \\
    \hline
    \multicolumn{4}{c}{NGC6266 (M62)} \\
    \hline
    2677 & 2002-05-12T09:12:42 & S & 62.27 \\
    15761 & 2014-05-05T19:18:39 & S & 82.09 \\
    \hline
    \multicolumn{4}{c}{NGC6440} \\
    \hline
    947 & 2000-07-04T13:28:39 & S & 23.28 \\
    3799 & 2003-06-27T08:57:31 & S & 24.05 \\
    10060 & 2009-07-28T15:05:44 & S & 49.11 \\
    \hline
    \multicolumn{4}{c}{NGC6441} \\
    \hline
    9598 & 2008-06-22T22:06:39 & S & 17.97 \\
    9874 & 2008-06-24T08:51:51 & S & 16.96 \\
    \hline
    \multicolumn{4}{c}{NGC6517} \\
    \hline
    9597 & 2009-02-04T14:22:23 & S & 23.62 \\
    \hline
    \multicolumn{4}{c}{NGC6544} \\
    \hline
    5435 & 2005-07-20T14:59:58 & S & 16.28 \\
    \hline
    \multicolumn{4}{c}{NGC6652} \\
    \hline
    12461 & 2011-06-03T21:35:28 & S & 45.32 \\
    18987 & 2017-05-22T08:13:22 & S & 10.04 \\
    \hline
    \multicolumn{4}{c}{NGC6656 (M22)} \\
    \hline
    5437 & 2005-05-24T21:21:23 & S & 15.82 \\
    14609 & 2014-05-22T19:39:17 & S & 84.86 \\
    \hline
    \multicolumn{4}{c}{NGC6760} \\
    \hline
    13672 & 2012-06-27T17:09:21 & S & 51.38 \\
    \hline
    \multicolumn{4}{c}{NGC6838 (M71)} \\
    \hline
    5434 & 2004-12-20T15:18:45 & S & 52.45 \\
    \hline
    \multicolumn{4}{c}{NGC7099 (M30)} \\
    \hline
    2679 & 2001-11-19T02:55:12 & S & 49.43 \\
    20725 & 2017-09-04T16:33:05 & S & 17.49 \\
    18997 & 2017-09-06T00:05:19 & S & 90.19 \\
    20726 & 2017-09-10T02:09:13 & S & 19.21 \\
    20732 & 2017-09-14T14:23:17 & S & 47.90 \\
    20731 & 2017-09-16T18:04:17 & S & 23.99 \\
    20792 & 2017-09-18T04:21:43 & S & 36.86 \\
    20795 & 2017-09-22T11:39:56 & S & 14.33 \\
    20796 & 2017-09-23T06:09:30 & S & 30.68 \\
    \hline
    \hline
    \end{tabular}
    \hfill
    \centering
    \begin{tabular}{c c c c}
    \hline
    \hline
    Obs.ID & Start Date and Time & Inst. & Exp. time \\
    & (UTC) & (ACIS-) & (ks) \\
    \hline
    \multicolumn{4}{c}{NGC104 (47Tuc)} \\
    \hline
    953 & 2000-03-16T08:38:40 & I & 31.67 \\
    955 & 2000-03-16T18:32:00 & I & 31.67 \\
    2735 & 2002-09-29T16:57:56 & S & 65.24 \\
    2736 & 2002-09-30T13:24:28 & S & 65.24 \\
    2737 & 2002-10-02T18:50:07 & S & 65.24 \\
    2738 & 2002-10-11T01:41:55 & S & 68.77 \\
    16527 & 2014-09-05T04:38:37 & S & 40.88 \\
    15747 & 2014-09-09T19:32:57 & S & 50.04 \\
    16529 & 2014-09-21T07:55:51 & S & 24.70 \\
    15748 & 2014-10-02T06:17:00 & S & 16.24 \\
    16528 & 2015-02-02T14:23:34 & S & 40.28 \\
    \hline
    \multicolumn{4}{c}{NGC6397} \\
    \hline
    79 & 2000-07-31T15:30:29 & I & 48.34 \\
    2668 & 2002-05-13T19:17:40 & S & 28.10 \\
    2669 & 2002-05-15T18:53:27 & S & 26.66 \\
    7461 & 2007-06-22T21:44:15 & S & 88.90 \\
    7460 & 2007-07-16T06:21:36 & S & 147.71 \\
    \hline
    \multicolumn{4}{c}{Terzan5} \\
    \hline
    3798 & 2003-07-13T13:22:45 & S & 39.34 \\
    10059 & 2009-07-15T17:19:56 & S & 36.26 \\
    13225 & 2011-02-17T09:05:34 & S & 29.67 \\
    13252 & 2011-04-29T17:06:31 & S & 39.54 \\
    13705 & 2011-09-05T16:54:24 & S & 13.87 \\
    14339 & 2011-09-08T03:32:23 & S & 34.06 \\
    13706 & 2012-05-13T17:58:45 & S & 46.46 \\
    14475 & 2012-09-17T16:10:24 & S & 30.50 \\
    14476 & 2012-10-28T03:14:38 & S & 28.60 \\ 
    14477 & 2013-02-05T04:16:59 & S & 28.60 \\
    14625 & 2013-02-22T08:22:32 & S & 49.20 \\
    15615 & 2013-02-23T10:17:02 & S & 84.16 \\
    14478 & 2013-07-16T21:12:59 & S & 28.60 \\
    14479 & 2014-07-15T05:23:11 & S & 28.60 \\
    16638 & 2014-07-17T11:48:31 & S & 71.60 \\
    15750 & 2014-07-20T16:41:37 & S & 22.99 \\
    17779 & 2016-07-13T18:41:43 & S & 68.85 \\
    18881 & 2016-07-15T11:50:35 & S & 64.71 \\
    \hline
    \multicolumn{4}{c}{NGC6626 (M28)} \\
    \hline
    2684 & 2002-07-04T18:02:19 & S & 12.75 \\
    2685 & 2002-08-04T23:46:25 & S & 13.51 \\
    2683 & 2002-09-09T16:55:03 & S & 14.11 \\
    9132 & 2008-08-07T20:45:43 & S & 142.26 \\
    9133 & 2008-08-10T23:50:24 & S & 54.46 \\
    14616 & 2013-04-28T19:37:19 & S & 14.79 \\
    16748 & 2015-05-30T02:34:33 & S & 29.66 \\
    16749 & 2015-08-07T20:13:25 & S & 29.55 \\
    16750 & 2015-11-07T16:05:40 & S & 29.57 \\
    \hline
    \multicolumn{4}{c}{NGC6752} \\
    \hline
    948 & 2000-05-15T04:36:02 & S & 29.47 \\
    6612 & 2006-02-10T22:48:48 & S & 37.97 \\
    19014 & 2017-07-02T03:27:25 & S & 98.81 \\
    19013 & 2017-07-24T09:33:12 & S & 43.20 \\
    20121 & 2017-07-25T17:04:15 & S & 18.26 \\
    20122 & 2017-07-29T09:00:43 & S & 67.22 \\
    20123 & 2017-07-30T23:53:18 & S & 49.46 \\
    \hline
    \hline
    \end{tabular}
    \end{center}
\textbf{Note.}
Only observations with the effective exposure $>10$~ks are used in our analysis. Data with shorter exposure (e.g. those were acquired from TOO/DDT requests) are excluded because either they do not provide sufficient photon statistics or they are affected by nearby outbursts.

\end{table}

\begin{table}
\begin{center}
\caption{\label{tab:Table2}Properties of 56 X-ray detected MSPs in GCs.}
\resizebox{\textwidth}{!}{
\tiny
\begin{tabular}{c c c c c c c c c c c}
\hline
\hline
MSP Name & R.A. (J2000) & Dec. (J2000) & $P$ & Class$^{a}$ & $P_{b}$ & Thermal (T) / & $\Gamma$ & log$_{10}L_{x}^{0.3-8}$ & log$_{10}L_{x}^{2-10}$ & Ref. \\
 & (h : m : s)& ($^{\circ}$ : $'$ : $''$) & (ms) & & (day) & Non-thermal (NT) & & (erg/s) & (erg/s) & \\
\hline
\multicolumn{11}{c}{NGC104 (47Tuc) / $d$ = 4.5 kpc \citep{Harris1996.112}} \\
\hline
47Tuc~aa & 00:24:07.31 & -72:05:19.40 & 1.840 & Iso & $\cdot\cdot$ & T & 2.13$^{+0.39}_{-0.36}$ & 30.35$^{+0.10}_{-0.09}$ & 29.98$^{+0.24}_{-0.27}$ & 1,2,3$^{*}$ \\
47Tuc~ab & 00:24:08.10 & -72:04:47.88 & 3.705 & Iso & $\cdot\cdot$ & T & 2.43$^{+0.17}_{-0.16}$ & 30.81$^{+0.03}_{-0.04}$ & 30.28$\pm0.11$ & 2,3$^{*}$,4 \\
47Tuc~C & 00:23.50:36 & -72:04:31.51 & 5.767 & Iso & $\cdot\cdot$ & T & 2.89$^{+0.41}_{-0.39}$ & 30.35$\pm0.08$ & 29.52$^{+0.26}_{-0.28}$ & 2,4,5$^{*}$ \\
47Tuc~D & 00:24:13.88 & -72:04:43.85 & 5.358 & Iso & $\cdot\cdot$ & T & 2.44$^{+0.21}_{-0.20}$ & 30.74$\pm0.04$ & 30.20$\pm0.14$ & 2,4,5$^{*}$ \\
47Tuc~E & 00:24:11.11 & -72:05:20.15 & 3.536 & Oth & 2.257 & T & 3.01$^{+0.20}_{-0.19}$ & 30.88$\pm0.04$ & 29.97$^{+0.13}_{-0.14}$ & 2,4,5$^{*}$ \\
47Tuc~H & 00:24:06.70 & -72:04:06.81 & 3.210 & Oth & 2.358 & T & 3.15$^{+0.28}_{-0.27}$ & 30.64$\pm0.06$ & 29.63$\pm0.18$ & 2,4,5$^{*}$ \\
47Tuc~J & 00:23:59.41 & -72:03:58.79 & 2.101 & BW & 0.127 & NT & 2.05$^{+0.16}_{-0.15}$ & 30.92$\pm0.04$ & 30.58$\pm0.10$ & 2,4,5$^{*}$ \\
47Tuc~L & 00:24:03.77 & -72:04:56.92 & 4.346 & Iso & $\cdot\cdot$ & T & 2.66$\pm0.11$ & 31.13$\pm0.02$ & 30.45$\pm0.07$ & 2,4,5$^{*}$ \\
47Tuc~M & 00:23:54.49 & -72:05:30.76 & 3.677 & Iso & $\cdot\cdot$ & T & 2.55$^{+0.31}_{-0.29}$ & 30.53$\pm0.07$ & 29.92$^{+0.19}_{-0.20}$ & 2,4,5$^{*}$ \\
47Tuc~N & 00:24:09.19 & -72:04:28.89 & 3.054 & Iso & $\cdot\cdot$ & T & 2.47$\pm0.24$ & 30.60$\pm0.05$ & 30.05$\pm0.16$ & 2,4,5$^{*}$ \\
47Tuc~O & 00:24:04.65 & -72:04:53.77 & 2.643 & BW & 0.136 & NT & 2.62$\pm0.12$ & 31.01$\pm0.03$ & 30.35$\pm0.08$ & 2,4,5$^{*}$ \\
47Tuc~Q & 00:24:16.49 & -72:04:25.16 & 4.033 & Oth & 1.189 & T & 2.63$^{+0.26}_{-0.25}$ & 30.62$\pm0.05$ & 29.96$^{+0.16}_{-0.17}$ & 2,4,5$^{*}$ \\
47Tuc~R & 00:24:07.59 & -72:04:50.40 & 3.480 & BW & 0.066 & T & 2.61$\pm0.15$ & 30.93$\pm0.03$ & 30.28$\pm0.10$ & 2,4,5$^{*}$ \\
47Tuc~T & 00:24:08.55 & -72:04:38.93 & 7.588 & Oth & 1.126 & T & 2.49$^{+0.32}_{-0.31}$ & 30.50$\pm0.06$ & 29.93$^{+0.20}_{-0.21}$ & 2,4,5$^{*}$ \\
47Tuc~U & 00:24:09.84 & -72:03:59.69 & 4.343 & Oth & 0.429 & T & 2.41$^{+0.24}_{-0.23}$ & 30.73$\pm0.05$ & 30.21$^{+0.14}_{-0.15}$ & 2,4,5$^{*}$ \\
47Tuc~W & 00:24:06.06 & -72:04:49.03 & 2.352 & RB & 0.133 & NT & 1.63$\pm0.07$ & 31.35$\pm0.02$ & 31.19$\pm0.04$ & 2,5$^{*}$,6 \\
47Tuc~X & 00:24:22.42 & -72:01:17.29 & 4.772 & Oth & 10.921 & T & 2.94$^{+0.44}_{-0.41}$ & 30.51$\pm0.09$ & 29.65$^{+0.25}_{-0.26}$ & 2,6$^{*}$ \\
47Tuc~Y & 00:24:01.40 & -72:04:41.84 & 2.197 & Oth & 0.522 & T & 3.04$^{+0.25}_{-0.24}$ & 30.65$\pm0.05$ & 29.72$^{+0.15}_{-0.16}$ & 2,4,5$^{*}$ \\
47Tuc~Z & 00:24:06.02 & -72:05:01.65 & 4.554 & Iso & $\cdot\cdot$ & T & 2.63$^{+0.17}_{-0.16}$ & 30.86$\pm0.03$ & 30.20$\pm0.11$ & 2,3$^{*}$,4 \\
\hline
\multicolumn{11}{c}{NGC 5139 ($\omega$ Cen) / $d$ = 5.2 kpc \citep{Harris1996.112}} \\
\hline
$\omega$ Cen~A & 13:26:39.67 & -47:30:11.64 & 4.109 & Iso & $\cdot\cdot$ & $\cdot\cdot$ & 2.87$^{+1.47}_{-1.32}$ & 30.31$^{+0.45}_{-0.25}$ & 29.50$^{+0.64}_{-0.72}$ & 2,7,8$^{*}$ \\
$\omega$ Cen~B & 13:26:49.57 & -47:29:24.18 & 4.792 & BW & 0.090 & $\cdot\cdot$ & 1.94$^{+0.69}_{-0.63}$ & 30.75$^{+0.14}_{-0.15}$ & 30.47$^{+0.33}_{-0.36}$ & 2,7,8$^{*}$,9 \\
\hline
\multicolumn{11}{c}{NGC6121 (M4) / $d$ = 1.73 kpc \citep{Richer1997.484}} \\
\hline
M4~A & 16:23:38.21 & -26:31:54.21 & 11.076 & Oth & 191.443 & $\cdot\cdot$ & 3.03$^{+0.32}_{-0.30}$ & 30.47$^{+0.10}_{-0.09}$ & 29.54$\pm0.16$ & 2,10,11$^{*}$ \\
\hline
\multicolumn{11}{c}{NGC6205 (M13) / $d$ = 7.1 kpc \citep{Harris1996.112}} \\
\hline
M13~B & 16:41:40.39 & 36:25:58.49 & 3.528 & Oth & 1.259 & NT & 1.80$\pm0.70$ & 30.90$^{+0.18}_{-0.30}$ & 30.70$^{+0.18}_{-0.30}$ & 12,13$^{*}$ \\
M13~C & 16:41:41.01 & 36:27:02.74 & 3.722 & Iso & $\cdot\cdot$ & T & 3.90$\pm1.20$ & 30.73$^{+0.22}_{-0.48}$ & 29.15$^{+0.22}_{-0.48}$ & 12,13$^{*}$ \\
M13~D & 16:41:42.40 & 36:27:28.20 & 3.118 & Oth & 0.591 & T & 3.70$\pm0.90$ & 30.89$^{+0.19}_{-0.34}$ & 29.74$^{+0.24}_{-0.30}$ & 12,13$^{*}$ \\
M13~E & 16:41:42.02 & 36:27:34.97 & 2.487 & BW & 0.113 & NT & 2.20$\pm0.60$ & 31.06$^{+0.14}_{-0.20}$ & 30.57$^{+0.13}_{-0.15}$ & 12,13$^{*}$ \\
M13~F & 16:41:44.61 & 36:28:16.00 & 3.004 & Oth & 1.378 & T & 3.70$\pm0.70$ & 31.08$^{+0.15}_{-0.22}$ & 29.75$^{+0.12}_{-0.13}$ & 12,13$^{*}$ \\
\hline
\multicolumn{11}{c}{NGC6266 (M62) / $d$ = 6.8 kpc \citep{Harris1996.112}} \\
\hline
M62~B & 17:01:12.70 & -30:06:48.87 & 3.594 & RB & 0.145 & NT & 2.17$^{+0.21}_{-0.20}$ & 31.95$\pm0.04$ & 31.56$^{+0.09}_{-0.10}$ & 14$^{*}$ \\
M62~C & 17:01:12.92 & -30:06:59.09 & 7.613 & Oth & 0.215 & NT & 2.42$\pm0.39$ & 31.65$^{+0.08}_{-0.06}$ & 31.12$\pm0.19$ & 14$^{*}$ \\
\hline
\multicolumn{11}{c}{NGC6397 / $d$ = 2.4 kpc \citep{Strickler2009.699}}\\
\hline
NGC6397~A & 17:40:44.57 & -53:40:41.84 & 3.650 & RB & 1.354 & NT & 1.85$\pm0.07$ & 31.35$^{+0.01}_{-0.02}$ & 31.11$\pm0.03$ & 15$^{*}$ \\
\hline
\multicolumn{11}{c}{Terzan5 / $d$ = 5.9 kpc \citep{Valenti2007.133}} \\
\hline
Terzan5~A & 17:48:02.25 & -24:46:37.91 & 11.563 & RB & 0.076 & NT & 1.24$\pm0.69$ & 31.95$^{+0.12}_{-0.16}$ & 31.91$^{+0.12}_{-0.16}$ & 16$^{*}$ \\
Terzan5~E & 17:48:03.41 & -24:46:35.78 & 2.198 & Oth & 60.060 & $\cdot\cdot$ & 2.00$^{b}$ & 30.49$\pm0.14$ & 30.18$\pm0.14$ & 16$^{*}$ \\
Terzan5~F & 17:48:05.11 & -24:46:38.16 & 5.554 & Iso & $\cdot\cdot$ & $\cdot\cdot$ & 2.00$^{b}$ & 30.95$\pm0.12$ & 30.64$\pm0.12$ & 16$^{*}$ \\
Terzan5~H & 17:48:05.62 & -24:46:53.24 & 4.926 & Iso & $\cdot\cdot$ & $\cdot\cdot$ & 2.00$^{b}$ & 30.86$\pm0.14$ & 30.55$\pm0.14$ & 16$^{*}$ \\
Terzan5~K & 17:48:03.90 & -24:46:47.84 & 2.970 & Iso & $\cdot\cdot$ & $\cdot\cdot$ & 2.00$^{b}$ & 30.35$\pm0.15$ & 30.04$\pm0.15$ & 16$^{*}$ \\
Terzan5~L & 17:48:04.74 & -24:46:35.75 & 2.245 & Iso & $\cdot\cdot$ & $\cdot\cdot$ & 2.00$^{b}$ & 31.15$\pm0.09$ & 30.84$\pm0.09$ & 16$^{*}$ \\
Terzan5~N & 17:48:04.91 & -24:46:54.00 & 8.667 & Oth & 0.386 &  $\cdot\cdot$ & 2.00$^{b}$ & 30.40$\pm0.13$ & 30.09$\pm0.13$ & 16$^{*}$ \\
Terzan5~O & 17:48:04.69 & -24:46:51.41 & 1.677 & BW & 0.260 & $\cdot\cdot$ & 2.00$^{b}$ & 31.46$\pm0.04$ & 31.15$\pm0.04$ & 16$^{*}$ \\
Terzan5~P & 17:48:05.04 & -24:46:41.40 & 1.396 & RB & 0.363 & NT & 0.86$^{+0.17}_{-0.16}$ & 32.59$\pm0.03$ & 32.28$\pm0.03$ & 16$^{*}$ \\
Terzan5~Q & 17:48:04.33 & -24:47:05.12 & 2.812 & Oth & 30.295 & $\cdot\cdot$ & 2.00$^{b}$ & 30.11$\pm0.19$ & 29.81$\pm0.17$ & 16$^{*}$ \\
Terzan5~V & 17:48:05.09 & -24:46:34.63 & 2.073 & Oth & 0.504 & $\cdot\cdot$ & 2.00$^{b}$ & 31.29$\pm0.06$ & 30.98$\pm0.06$ & 16$^{*}$ \\
Terzan5~X & 17:48:05.59 & -24:47:12.14 & 2.999 & Oth & 4.999 & $\cdot\cdot$ & 2.00$^{b}$ & 30.71$\pm0.15$ & 30.40$\pm0.15$ & 16$^{*}$ \\
Terzan5~Z & 17:48:04.95 & -24:46:46.04 & 2.463 & Oth & 3.488 & $\cdot\cdot$ & 2.00$^{b}$ & 31.06$\pm0.07$ & 30.76$\pm0.07$ & 16$^{*}$ \\
Terzan5~ad & 17:48:03.85 & -24:46:41.94 & 1.729 & RB & 1.094 & NT & 1.16$^{+0.24}_{-0.22}$ & 32.30$\pm0.04$ & 32.89$\pm0.04$ & 16$^{*}$ \\
\hline
\multicolumn{11}{c}{NGC6626 (M28) / $d$ = 5.5 kpc \citep{Harris1996.112}} \\
\hline
M28~A & 18:24:32.01 & -24:52:10.83 & 3.054 & Iso & $\cdot\cdot$ & NT & 1.21$\pm0.02$ & 33.15$\pm0.01$ & 33.12$\pm0.01$ & 2,17$^{*}$ \\
M28~G & 18:24:33.03 & -24:52:17.32 & 5.909 & BW & 0.105 & NT & 2.89$\pm0.25$ & 31.29$\pm0.07$ & 30.47$^{+0.12}_{-0.13}$ & 2,17$^{*}$ \\
M28~H & 18:24:31.61 & -24:52:17.20 & 4.629 & RB & 0.435 & NT & 1.11$\pm0.19$ & 31.34$^{+0.06}_{-0.05}$ & 31.33$\pm0.09$ & 2,17$^{*}$,18 \\
M28~I & 18:24:32.53 & -24:52:08.60 & 3.932 & RB & 0.459 & NT & 1.46$\pm0.28$ & 32.46$\pm0.68$ & 32.36$\pm0.68$ & 2,19$^{*}$,20 \\
M28~L & 18:24:32.35 & -24:52:08.02 & 4.100 & BW & 0.226 & NT & 1.40$\pm0.06$ & 32.27$\pm0.02$ & 32.19$\pm0.03$ & 2,17$^{*}$,21 \\
\hline
\multicolumn{11}{c}{NGC6656 (M22) / $d$ = 3.2 kpc \citep{Harris1996.112}} \\
\hline
M22~A & 18:36:25.50 & -23:54:51.50 & 3.350 & BW & 0.203 & $\cdot\cdot$ & 1.50$^{+0.70}_{-0.60}$ & 30.47$\pm0.12$ & 30.38$\pm0.12$ & 22$^{*}$ \\
\hline
\multicolumn{11}{c}{NGC6752 / $d$ = 4.0 kpc \citep{Harris1996.112}} \\
\hline
NGC6752~A & 19:11:42.76 & -59:58:26.90 & 3.266 & Oth & 0.837 & T & 3.06$^{+0.49}_{-0.47}$ & 30.86$^{+0.15}_{-0.14}$ & 29.91$^{+0.24}_{-0.26}$ & 2,23$^{*}$ \\
NGC6752~B & 19:10:52.05 & -59:59:00.75 & 8.358 & Iso & $\cdot\cdot$ & T & 3.40$^{+1.53}_{-1.45}$ & 30.76$^{+0.49}_{-0.27}$ & 29.56$^{+0.83}_{-0.78}$ & 2,23$^{*}$ \\
NGC6752~C & 19:11:05.56 & -60:00:59.70 & 5.277 & Iso & $\cdot\cdot$ & T & 2.51$^{+0.36}_{-0.34}$ & 30.69$\pm0.08$ & 30.11$^{+0.19}_{-0.20}$ & 2,23$^{*}$ \\
NGC6752~D & 19:10:52.42 & -59:59:05.50 & 9.034 & Iso & $\cdot\cdot$ & T & 2.60$\pm0.27$ & 30.92$\pm0.05$ & 30.28$\pm0.16$ & 2,23$^{*}$ \\
NGC6752~F & 19:10:52.07 & -59:59:09.41 & 4.143 & Iso & $\cdot\cdot$ & $\cdot\cdot$ & 2.30$\pm0.65$ & 30.93$\pm0.20$ & 30.47$\pm0.20$ & 2,24,25$^{*}$ \\
\hline
\multicolumn{11}{c}{NGC6838 (M71) / $d$ = 4.0 kpc \citep{Harris1996.112}} \\
\hline
M71~A & 19:53:46.42 & 18:47:04.91 & 4.888 & BW & 0.177 & NT & 1.85$^{+0.33}_{-0.31}$ & 31.19$\pm0.07$ & 30.95$\pm0.16$ & 2,26$^{*}$ \\
\hline
\hline
\end{tabular}}
\end{center}
\scriptsize
\textbf{Note.}

$^a$ Different classes of MSPs: Iso (Isolated MSP), RB (Redback), BW (Black-widow), and Oth (Others binary). 

$^b$ Following \cite{Bogdanov2021.912}, a single power-law model with $\Gamma$ fixed at 2 are adopted for these cases because of the small photon statistics.

\textbf{Ref.} (1) \cite{Freire2018.476} (2) this work (3) \cite{Bhattacharya2017.472} (4) \cite{Freire2017.471} (5) \cite{Bogdanov2006.646} \ 
(6) \cite{Ridolfi2016.462} (7) \cite{Dai2020.888} (8) \cite{Zhao2022MNRAS.511.5964Z} (9) \cite{Henleywillis2018MNRAS.479} (10) \cite{Bassa2004.609} \ 
(11) \cite{Pavlov2007.664} (12) \cite{Wang2020.892} (13) \cite{Zhao2021.502} (14) \cite{Oh2020.498} (15) \cite{Bogdanov2010.709} \ 
(16) \cite{Bogdanov2021.912} (17) \cite{Bogdanov2011.730} (18) \cite{Pallanca2010ApJ.725} (19) \cite{Papitto2013.501} (20) \cite{Linares2014.438} \ 
(21) \cite{Becker2003ApJ.594} (22) \cite{Amato2019.486} (23) \cite{Forestell2014.441} (24) \cite{Ridolfi.2021.MNRAS.504} (25) \cite{Cohn2021MNRAS.508.2823C} \ 
(26) \cite{Elsner2008.687} \
The entries marked with * are the references which have the identifications of the X-ray counterparts of GC MSPs reported.
\end{table}

\begin{table}
\begin{center}
\caption{\label{tab:Table3}Upper limits of $L_{x}$ of 60 GC MSPs.}
\tiny
\begin{tabular}{c c c c c c c c c}
\hline
\hline
MSP Name & R.A. (J2000) & Dec. (J2000) & $P$ & Class & $P_{b}$ & log$_{10}L_{x}^{0.3-8}$ & log$_{10}L_{x}^{2-10}$ & Ref.\\
 & (h : m : s)& ($^{\circ}$ : $'$ : $''$) & (ms) &  & (day) & (erg/s) & (erg/s) & \\
\hline
\multicolumn{8}{c}{NGC104 (47Tuc) / $d$ = 4.5 kpc \citep{Harris1996.112}} \\
\hline
47Tuc~F & 00:24:03.86 & -72:04:42.82 & 2.624 & Iso & $\cdot\cdot$ & $\leq$30.01 & $\leq$29.64 & 1 \\
47Tuc~G & 00:24:07.96 & -72:04:39.70 & 4.040 & Iso & $\cdot\cdot$ & $\leq$30.01 & $\leq$29.64 & 1 \\
47Tuc~I & 00:24:07.93 & -72:04:39.68 & 3.485 & BW & 0.230 & $\leq$30.01 & $\leq$29.64 & 1 \\
47Tuc~P & 00:24:2s0 & -72:04:10 & 3.643 & BW & 0.147 & $\leq$30.01 & $\leq$29.64 & 2 \\
47Tuc~S & 00:24:03.98 & -72:04:42.35 & 2.830 & Oth & 1.201 & $\leq$30.01 & $\leq$29.64 & 1 \\
47Tuc~V & 00:24:05.36 & -72:04:53.20 & 4.810 & Oth & 0.227 & $\leq$30.01 & $\leq$29.64 & 2 \\
\hline
\multicolumn{8}{c}{NGC1851 / $d$ = 12.1 kpc \citep{Harris1996.112}} \\
\hline
NGC1851~A & 05:14:06.69 & -40:02:48.89 & 4.991 & Oth & 18.785 & $\leq$31.96 & $\leq$31.65 & 3 \\
\hline
\multicolumn{8}{c}{NGC5272 (M3) / $d$ = 10.2 kpc \citep{Harris1996.112}} \\
\hline
M3~B & 13:42:11.09 & 28:22:40.14 & 2.389 & Oth & 1.417 & $\leq$32.82 & $\leq$31.81 & 4 \\
M3~D & 13:42:10.20 & 28:22:36.00 & 5.443 & Oth & 128.752 & $\leq$32.82 & $\leq$31.81 & 4 \\
\hline
\multicolumn{8}{c}{NGC5904 (M5) / $d$ = 7.5 kpc\citep{Harris1996.112}} \\
\hline
M5~A & 15:18:33.32 & 02:05:27.44 & 5.554 & Iso & $\cdot\cdot$ & $\leq$31.41 & $\leq$31.04 & 5 \\
M5~B & 15:18:31.46 & 02:05:15.30 & 7.947 & Oth & 6.859 & $\leq$31.41 & $\leq$31.04 & 5 \\
M5~C & 15:18:32.79 & 02:04:47.82 & 2.484 & BW & 0.087 & $\leq$31.41 & $\leq$31.04 & 6 \\
\hline
\multicolumn{8}{c}{NGC6205 (M13) / $d$ = 7.1 kpc \citep{Harris1996.112}} \\
\hline
M13~A & 16:41:40.87 & 36:27:14.98 & 10.38 & Iso & $\cdot\cdot$ & $\leq$31.43 & $\leq$31.12 & 7 \\
\hline
\multicolumn{8}{c}{NGC6266 (M62) / $d$ = 6.9 kpc \citep{Harris1996.112}} \\
\hline
M62~A & 17:01:12.51 & -30:06:30.17 & 5.242 & Oth & 3.806 & $\leq$31.04 & $\leq$30.73 & 8 \\
M62~D & 17:01:13.56 & -30:06:42.56 & 3.418 & Oth & 1.118 & $\leq$31.04 & $\leq$30.73 & 8 \\
M62~E & 17:01:13.27 & -30:06:46.89 & 3.234 & BW & 0.159 & $\leq$31.04 & $\leq$30.73 & 8 \\
M62~F & 17:01:12.82 & -30:06:51.82 & 2.295 & BW & 0.206 & $\leq$31.04 & $\leq$30.73 & 8 \\
M62~G & 17:01:14.00 & -30:06:42.00 & 4.608 & Oth & 0.774 & $\leq$31.04 & $\leq$30.73 & 9 \\
\hline
\multicolumn{8}{c}{NGC6440 / $d$ = 8.5 kpc \citep{Harris1996.112}} \\
\hline
NGC6440~B & 17:48:52.76 & -20:21:38.45 & 16.760 & Oth & 20.550 & $\leq$31.72 & $\leq$31.41 & 10 \\
NGC6440~C & 17:48:51.17 & -20:21:53.81 & 6.227 & Iso & $\cdot\cdot$ & $\leq$31.72 & $\leq$31.41 & 11 \\
NGC6440~D & 17:48:51.65 & -20:21:07.41 & 13.496 & RB & 0.286 & $\leq$31.72 & $\leq$31.41 & 11 \\
NGC6440~E & 17:48:52.80 & -20:21:29.38 & 16.264 & Iso & $\cdot\cdot$ & $\leq$31.72 & $\leq$31.41 & 11 \\
NGC6440~F & 17:48:52.33 & -20:21:39.33 & 3.794 & Oth & 9.834 & $\leq$31.72 & $\leq$31.41 & 11 \\
\hline
\multicolumn{8}{c}{Terzan5 / $d$ = 5.9 kpc \citep{Valenti2007.133}} \\
\hline
Terzan5~C & 17:48:04.52 & -24:46:35.17 & 8.436 & Iso & $\cdot\cdot$ & $\leq$31.80 & $\leq$31.49 & 12 \\
Terzan5~D & 17:48:05.93 & -24:46:06.05 & 4.714 & Iso & $\cdot\cdot$ & $\leq$31.80 & $\leq$31.49 & 12 \\
Terzan5~I & 17:48:04.85 & -24:46:46.37 & 9.570 & Oth & 1.300 & $\leq$31.80 & $\leq$31.49 & 12 \\
Terzan5~M & 17:48:04.62 & -24:46:40.75 & 3.570 & Oth & 0.443 & $\leq$31.80 & $\leq$31.49 & 12 \\
Terzan5~R & 17:48:04.69 & -24:46:50.25 & 5.029 & Iso & $\cdot\cdot$ & $\leq$31.80 & $\leq$31.49 & 12 \\
Terzan5~S & 17:48:04.29 & -24:46:31.71 & 6.117 & Iso & $\cdot\cdot$ & $\leq$31.80 & $\leq$31.49 & 12 \\
Terzan5~T & 17:48:02.99 & -24:46:52.81 & 7.045 & Iso & $\cdot\cdot$ & $\leq$31.80 & $\leq$31.49 & 12 \\
Terzan5~U & 17:48:04.24 & -24:46:47.86 & 3.289 & Oth & 3.600 & $\leq$31.80 & $\leq$31.49 & 12 \\
Terzan5~W & 17:48:04.84 & -24:46:42.38 & 4.205 & Oth & 4.877 & $\leq$31.80 & $\leq$31.49 & 12 \\
Terzan5~Y & 17:48:05.10 & -24:46:44.57 & 2.048 & Oth & 1.170 & $\leq$31.80 & $\leq$31.49 & 12 \\
Terzan5~aa & 17:48:05.81 & -24:46:42.24 & 5.788 & Iso & $\cdot\cdot$ & $\leq$31.80 & $\leq$31.49 & 12 \\
Terzan5~ab & 17:48:04.76 & -24:46:42.65 & 5.120 & Iso & $\cdot\cdot$ & $\leq$31.80 & $\leq$31.49 & 12 \\
Terzan5~ac & 17:48:06.04 & -24:46:32.53 & 5.087 & Iso & $\cdot\cdot$ & $\leq$31.80 & $\leq$31.49 & 12 \\
Terzan5~ae & 17:48:04.96 & -24:46:45.72 & 3.659 & BW & 0.171 & $\leq$31.80 & $\leq$31.49 & 12 \\
Terzan5~af & 17:48:04.21 & -24:46:45.72 & 3.304 & Iso & $\cdot\cdot$ & $\leq$31.80 & $\leq$31.49 & 12 \\
Terzan5~ag & 17:48:04.81 & -24:46:34.59 & 4.448 & Iso & $\cdot\cdot$ & $\leq$31.80 & $\leq$31.49 & 12 \\
Terzan5~ah & 17:48:04.32 & -24:46:42.03 & 4.965 & Iso & $\cdot\cdot$ & $\leq$31.80 & $\leq$31.49 & 12 \\
Terzan5~aj & 17:48:05.01 & -24:46:34.69 & 2.959 & Iso & $\cdot\cdot$ & $\leq$31.80 & $\leq$31.49 & 12 \\
Terzan5~ak & 17:48:03.69 & -24:46:37.93 & 1.890 & Iso & $\cdot\cdot$ & $\leq$31.80 & $\leq$31.49 & 12 \\
\hline
\multicolumn{8}{c}{NGC6441 / $d$ = 11.6 kpc \citep{Harris1996.112}} \\
\hline
NGC6441~B & 17:50:12.18 & -37:03:22.93 & 6.075 & Oth & 3.605 & $\leq$32.18 & $\leq$31.80 & 11 \\
NGC6441~D & 17:50:13.10 & -37:03:06.37 & 5.140 & Iso & $\cdot\cdot$ & $\leq$32.18 & $\leq$31.80 & 11 \\
\hline
\multicolumn{8}{c}{NGC6517 / $d$ = 10.6 kpc \citep{Harris1996.112}} \\
\hline
NGC6517~A & 18:01:50.61 & -08:57:31.85 & 7.176 & Iso & $\cdot\cdot$ & $\leq$32.06 & $\leq$31.67 & 13 \\
NGC6517~C & 18:01:50.74 & -08:57:32.70 & 3.739 & Iso & $\cdot\cdot$ & $\leq$32.06 & $\leq$31.67 & 13 \\
NGC6517~D & 18:01:55.37 & -08:57:24.33 & 4.227 & Iso & $\cdot\cdot$ & $\leq$32.06 & $\leq$31.67 & 13 \\
\hline
\multicolumn{8}{c}{NGC6544 / $d$ = 3.0 kpc \citep{Harris1996.112}} \\
\hline
NGC6544~A & 18:07:20.36 & -24:59:52.90 & 3.059 & BW & 0.0711 & $\leq$31.14 & $\leq$30.75 & 8 \\
NGC6544~B & 18:07:20.87 & -25:00:01.92 & 4.186 & Oth & 9.957 & $\leq$31.14 & $\leq$30.75 & 8 \\
\hline
\multicolumn{8}{c}{NGC6626 (M28) / $d$ = 5.5 kpc \citep{Harris1996.112}} \\
\hline
M28~B & 18:24:32.55 & -24:52:04.29 & 6.547 & Iso & $\cdot\cdot$ & $\leq$30.83 & $\leq$30.52 & 14 \\
M28~C & 18:24:32.19 & -24:52:14.66 & 4.159 & Oth & 8.078 & $\leq$30.83 & $\leq$30.52 & 14 \\
M28~E & 18:24:33.09 & -24:52:13.57 & 5.420 & Iso & $\cdot\cdot$ & $\leq$30.83 & $\leq$30.52 & 14 \\
M28~F & 18:24:31.81 & -24:49:25.03 & 2.451 & Iso & $\cdot\cdot$ & $\leq$30.83 & $\leq$30.52 & 14 \\
M28~J & 18:24:32.73 & -24:52:10.18 & 4.039 & BW & 0.0974 & $\leq$30.83 & $\leq$30.52 & 14 \\
\hline
\multicolumn{8}{c}{NGC6652 / $d$ = 10.0 kpc \citep{Harris1996.112}} \\
\hline
NGC6652~A & 18:35:44.86 & -32:59:25.08 & 3.889 & Iso & $\cdot\cdot$ & $\leq$31.25 & $\leq$30.88 & 15 \\
\hline
\multicolumn{8}{c}{NGC6656 (M22) / $d$ = 3.2 kpc \citep{Harris1996.112}} \\
\hline
M22~B & 18:36:24.35 & -23:54:28.70 & 3.232 & Iso & $\cdot\cdot$ & $\leq$30.46 & $\leq$30.08 & 13 \\
\hline
\multicolumn{8}{c}{NGC6752 / $d$ = 4.0 kpc \citep{Harris1996.112}} \\
\hline
NGC6752~E & 19:10:52.16 & -59:59:02.10 & 4.572 & Iso & $\cdot\cdot$ & $\leq$29.94 & $\leq$29.87 & 16 \\
\hline
\multicolumn{8}{c}{NGC6760 / $d$ = 7.4 kpc \citep{Harris1996.112}} \\
\hline
NGC6760~A & 19:11:11.09 & 01:02:09.74 & 3.619 & BW & 0.1410 & $\leq$31.36 & $\leq$30.97 & 17 \\
NGC6760~B & 19:11:12.57 & 01:01:50.44 & 5.384 & Iso & $\cdot\cdot$ & $\leq$31.36 & $\leq$30.97 & 17 \\
\hline
\multicolumn{8}{c}{NGC7099 (M30) / $d$ = 8.1 kpc \citep{Harris1996.112}} \\
\hline
M30~A & 21:40:22.41 & -23:10:48.79 & 11.019 & RB & 0.1740 & $\leq$31.07 & $\leq$30.76 & 18 \\
\hline
\hline
\end{tabular}
\end{center}
\scriptsize
\textbf{Note.} Luminosities in column 7 \& 8 were calculated by assuming a single power-law model with a fixed $\Gamma$=2.

\textbf{Ref.} (1)\cite{Freire2017.471} (2) \cite{Ridolfi2016.462} (3) \cite{Ridolfi2019.490} (4) \cite{Cadelano2019.875} (5) \cite{Freire2008.679} \ 
(6) \cite{Pallanca2014.795} (7) \cite{Wang2020.892} (8) \cite{Lynch2012.745} (9) \cite{Ridolfi.2021.MNRAS.504} (10) \cite{Vleeschower2022MNRAS.513.1386V}\ 
(11) \cite{Freire2008.675} (12) \cite{Prager2017ApJ.845} (13) \cite{Lynch2011.734} (14) \cite{Bogdanov2011.730} (15) \cite{Decesar2015.807} \ 
(16) \cite{Forestell2014.441} (17) \cite{Freire2005.621} (18) \cite{Ransom2004.604}

\end{table}

\begin{table}
\begin{center}
\footnotesize
\caption{\label{tab:Table4} Derived parameters of 25 MSPs in GCs with estimates of intrinsic spin-down rate $\dot{P}_{\rm int}$}
\begin{tabular}{c c c c c c}
\hline
\hline
MSP Name & $P$ & $\dot{P}_{\textrm{int}}$ & log$_{10} \dot{E}$ & log$_{10} B_{s}$ & Ref.\\
 & (ms) & (s/s) & (erg s$^{-1}$) & (G) \\
\hline
\hline
\multicolumn{6}{c}{Correction by long-term radio timing} \\
\hline
47Tuc~E & 3.536 & (1.2$\pm0.4) \times 10^{-20}$ & 34.03$^{+0.12}_{-0.16}$ & 8.31$^{+0.06}_{-0.08}$ & 1 \\
47Tuc~H & 3.210 & (9.0$\pm9.0) \times 10^{-21}$ & 34.03$^{+0.30}_{-34.03}$ & 8.23$^{+0.15}_{-8.23}$ & 1 \\
47Tuc~J & 2.101 & (3.6$\pm2.3) \times 10^{-19}$ & 36.19$^{+0.22}_{-0.46}$ & 8.94$^{+0.11}_{-0.23}$ & 2 \\
47Tuc~R & 3.480 & (3.1$\pm2.2) \times 10^{-20}$ & 34.46$^{+0.23}_{-0.54}$ & 8.52$^{+0.12}_{-0.27}$ & 1 \\
47Tuc~T & 7.588 & (9.9$\pm8.9) \times 10^{-20}$ & 33.95$^{+0.28}_{-1.00}$ & 8.94$^{+0.14}_{-0.50}$ & 1 \\
47Tuc~U & 4.343 & (1.8$\pm0.5) \times 10^{-20}$ & 33.94$^{+0.11}_{-0.14}$ & 8.45$^{+0.05}_{-0.07}$ & 1 \\
47Tuc~Y & 2.197 & (4.7$\pm3.3) \times 10^{-21}$ & 34.24$^{+0.23}_{-0.53}$ & 8.01$^{+0.12}_{-0.26}$ & 1 \\
Terzan5~ae$^{+}$ & 3.659 & (3.9$\pm2.6) \times 10^{-19}$ & 35.50$^{+0.22}_{-0.46}$ & 9.08$^{+0.11}_{-0.23}$ & 2 \\
Terzan5~M$^{+}$ & 3.570 & (1.8$\pm1.1) \times 10^{-20}$ & 34.18$^{+0.22}_{-0.46}$ & 8.40$^{+0.11}_{-0.23}$ & 2 \\
Terzan5~N & 8.667 & (1.7$\pm1.1) \times 10^{-19}$ & 34.01$^{+0.22}_{-0.46}$ & 9.08$^{+0.11}_{-0.23}$ & 2 \\
Terzan5~O & 1.677 & (1.6$\pm1.0) \times 10^{-19}$ & 36.13$^{+0.22}_{-0.46}$ & 8.72$^{+0.11}_{-0.23}$ & 2 \\
Terzan5~V & 2.073 & (3.8$\pm2.5) \times 10^{-20}$ & 35.22$^{+0.22}_{-0.46}$ & 8.45$^{+0.11}_{-0.23}$ & 2 \\
Terzan5~W$^{+}$ & 4.205 & (1.1$\pm0.7)\times 10^{-19}$ & 34.78$^{+0.22}_{-0.46}$ & 8.84$^{+0.11}_{-0.23}$ & 2 \\
Terzan5~X & 2.999 & (2.6$\pm1.7) \times 10^{-20}$ & 34.58$^{+0.22}_{-0.46}$& 8.45$^{+0.11}_{-0.23}$  & 2 \\
Terzan5~Y$^{+}$ & 2.048 & (9.5$\pm6.1) \times 10^{-20}$ & 35.64$^{+0.22}_{-0.46}$ & 8.64$^{+0.11}_{-0.23}$ & 2 \\
Terzan5~Z & 2.463 & (1.2$\pm0.8) \times 10^{-20}$ & 34.49$^{+0.22}_{-0.46}$ & 8.23$^{+0.11}_{-0.23}$ & 2 \\
\hline
\multicolumn{6}{c}{Correction by King model} \\
\hline
47Tuc~C & 5.767 & (2.4$^{+5.8}_{-2.4}) \times 10^{-21}$ & 32.70$^{+0.53}_{-32.70}$ & 8.07$^{+0.27}_{-8.07}$ & 3 \\
47Tuc~D & 5.358 & (2.6$^{+1.1}_{-0.8}) \times 10^{-20}$ & 33.83$^{+0.15}_{-0.16}$ & 8.57$^{+0.07}_{-0.08}$ & 3 \\
47Tuc~L & 4.346 & (2.2$^{+3.7}_{-2.2}) \times 10^{-20}$ & 34.02$^{+0.43}_{-34.02}$ & 8.49$^{+0.22}_{-8.49}$ & 3 \\
47Tuc~M & 3.677 & (4.0$\pm3.0) \times 10^{-21}$ & 33.50$^{+0.24}_{-0.60}$ & 8.08$^{+0.12}_{-0.30}$ & 4 \\
47Tuc~N & 3.054 & (1.3$^{+1.1}_{-0.8}) \times 10^{-20}$ & 34.27$^{+0.25}_{-0.36}$ & 8.31$^{+0.13}_{-0.18}$ & 3 \\
47Tuc~O & 2.643 & (1.5$^{+0.4}_{-0.5}) \times 10^{-20}$ & 34.49$^{+0.10}_{-0.20}$ & 8.29$^{+0.05}_{-0.10}$ & 3 \\
47Tuc~Q & 4.033 & (3.0$\pm0.2) \times 10^{-20}$ & 34.26$\pm0.03$ & 8.54$\pm0.01$ & 3 \\
NGC6397~A & 3.650 & (1.6$\pm1.2) \times 10^{-19}$ & 35.10$^{+0.24}_{-0.60}$ & 8.88$^{+0.12}_{-0.30}$ & 4 \\
\hline
\multicolumn{6}{c}{ M28~A (MSP with large $\dot{P}$)} \\
\hline
M28~A & 3.054 & (1.62$\pm9\times10^{-7}) \times 10^{-18}$ & 36.35$\pm(2\times10^{-7})$ & 9.35$\pm(1\times10^{-7})$ & 5,6\\
\hline
\hline
\end{tabular}

\scriptsize
\textbf{Note.} The flag ``+'' in column 1 indicates the MSPs without X-ray counterparts identified (cf. \autoref{tab:Table3}). 

\textbf{Ref.} (1) \cite{Freire2017.471} (2) \cite{Prager2017ApJ.845} (3) \cite{Bogdanov2006.646} (4) \cite{Grindlay2002.581} / (5) \cite{Foster1988ApJ.326L.13F}/ (6) \cite{Cognard1996A&A.311.179C}
\end{center}
\end{table}

\begin{table}
\begin{center}
\caption{\label{tab:Table5} The results of Spearman rank test for the correlation between $L_{x}$ and $\dot{E}$ with the samples of GC MSPs in three different groups (see main text).}
\begin{tabular}{c | c c | c c}
\hline
\hline
 & \multicolumn{2}{c|}{${0.3-8}$ keV} & \multicolumn{2}{c}{${2-10}$ keV} \\
 \hline
 & Spearman'r & $p-$value & Spearman'r & $p-$value \\
\hline
Group A & 0.692 & 0.013 & 0.727 & 0.007 \\
Group B & 0.676 & 0.001 & 0.734 & 2$\times 10^{-4}$ \\
Group C & 0.721 & 2$\times 10^{-4}$ & 0.770 & 4$\times 10^{-5}$ \\
\hline
\hline
\end{tabular}
\end{center}
\end{table}

\begin{table}
\begin{center}
\caption{\label{tab:Table6} The results of Spearman rank test for the correlation between $P_{b}$ and $P$ with radio/X-ray selected MSPs in GCs and GF.}
\begin{tabular}{c | c c | c c}
\hline
\hline
 & \multicolumn{2}{c|}{Radio selected} & \multicolumn{2}{c}{X-ray selected} \\
 \hline
 & Spearman'r & $p-$value & Spearman'r & $p-$value \\
\hline
GC & 0.164 & 0.126 & -0.111 & 0.512 \\
GF & 0.391 & 7$\times 10^{-9}$ & 0.304 & 0.060 \\
\hline
\hline
\end{tabular}
\end{center}
\end{table}

\begin{table}
\begin{center}
\caption{\label{tab:Table7} Null hypothesis probabilities of K-S and A-D tests for comparing the X-ray properties between the MSPs in GCs and GF.}
\begin{tabular}{c | c c | c c | c c | c c | c c}
\hline
\hline
 & \multicolumn{2}{c |}{Whole} & \multicolumn{2}{c |}{Iso} & \multicolumn{2}{c |}{Oth} & \multicolumn{2}{c |}{BW} & \multicolumn{2}{c}{RB} \\
\hline
 & K-S & A-D & K-S & A-D & K-S & A-D & K-S & A-D & K-S & A-D \\
\hline
$\Gamma$ & 0.552 & 0.534 & 0.053 & 0.110 & 0.360 & 0.344 & 0.254 & 0.190 & 0.929 & 0.935 \\
$L_{x}^{2-10}$ & 0.033 & 0.019 & 0.032 & 0.077 & 0.161 & 0.062 & 0.227 & 0.227 & 0.519 & 0.433 \\
\hline
\hline
\end{tabular}
\end{center}
\end{table}

\begin{table}
\begin{center}
\caption{\label{tab:Table8} Null hypothesis probabilities of K-S and A-D tests for comparing the physical properties between radio/X-ray selected MSPs in GCs and GF. Both results with the GC MSP samples adopted from Group B and Group C are given in this table.}
\begin{tabular}{c | c c c c | c c c c}
\hline
\hline
 & \multicolumn{4}{c |}{GC vs. GF} & \multicolumn{4}{c}{Radio selected vs. X-ray selected} \\
\hline
 & \multicolumn{2}{c}{Radio selected} & \multicolumn{2}{c |}{X-ray selected} & \multicolumn{2}{c}{GC} & \multicolumn{2}{c}{GF} \\
\hline
 & K-S & A-D & K-S & A-D & K-S & A-D & K-S & A-D \\
\hline
$P_{b}$ & 5$\times 10^{-7}$ & 2$\times 10^{-5}$ & 0.464 & 0.748 & 0.722 & 0.710 & 2$\times$10$^{-4}$ & 0.001 \\
$P$ & 0.036 & 0.007 & 0.013 & 0.016 & 0.241 & 0.125 & 8$\times$10$^{-4}$ & 0.002 \\
$B_{s}$ (outliers excluded) & 0.005 & 0.002 & 0.019 & 0.019 & 1.0 & 1.0 & 0.726 & 0.622 \\
$B_{s}$ (outliers included) & 0.004 & 0.001 & 0.022 & 0.019 & 1.0 & 1.0 & 0.736 & 0.673 \\
$\dot{E}$ (outliers excluded) & 5$\times$10$^{-5}$ & 2$\times$10$^{-4}$ & 0.807 & 0.711 & 1.0 & 1.0 & 1$\times$10$^{-5}$ & 1$\times$10$^{-4}$ \\
$\dot{E}$ (outliers included) & 5$\times 10^{-5}$ & 2$\times$10$^{-4}$ & 0.715 & 0.557 & 1.0 & 1.0 & 7$\times$10$^{-6}$ & 9$\times$10$^{-5}$ \\
\hline
\hline
\end{tabular}
\end{center}
\end{table}

\begin{table}
\begin{center}
\caption{\label{tab:Table9} The median for the physical properties between radio/X-ray selected MSPs in GCs and GF. For $B_{s}$ and $\dot{E}$, the values inside and outside the parentheses are the medians with the outliers excluded and included respectively.}
\begin{tabular}{c | c c | c c}
\hline
\hline
 & \multicolumn{2}{c |}{GF} & \multicolumn{2}{c}{GC} \\
\hline
& Radio selected & X-ray selected & Radio selected & X-ray selected \\
\hline
$P_{b}$ (day) & 4.78 & 0.61 & 0.79 & 0.46 \\
$P$ (ms) & 3.73 & 2.95 & 4.34 & 3.69 \\
$B_{s}$ ($10^{8}$~G) & 1.94 (1.93) & 1.91 (1.82) & 3.07 (2.95) & 2.83 (2.83) \\
$\dot{E}$ ($10^{34}$~erg/s) & 0.49 (0.49) & 1.73 (1.63) & 1.87 (1.85) & 1.82 (1.79) \\
\hline
\hline
\end{tabular}
\end{center}
\end{table}

\begin{table}
\begin{center}
\caption{\label{tab:Table10} Parameters for various assumed linear models as inferred from Bayesian regression. The best-fit values correspond to the peak values of the marginalized posterior probability distributions. And the uncertainties are estimated from the ranges with 68\% of the samples bracketed. }
\begin{tabular}{c | c c | c c | c c | c c}
\hline
\hline
 & Slope & Intercept & Slope & Intercept & Slope & Intercept & Slope & Intercept \\
\hline
 & \multicolumn{2}{c |}{GF (outliers excluded)} & \multicolumn{2}{c |}{GF (All)} & \multicolumn{2}{c |}{GC (Group B)} & \multicolumn{2}{c}{GC (Group C)}\\
\hline
$L_{x}^{0.3-8}-\dot{E}$ & 0.86$\pm$0.16 & 1.36$\pm$5.36 & 0.94$\pm$0.12 & -1.30$\pm$4.17 & 0.36$\pm$0.09 & 18.51$\pm$3.19 & 0.56$\pm$0.14 & 11.43$\pm$4.90 \\
$L_{x}^{2-10}-\dot{E}$ & 1.03$\pm$0.26 & -4.92$\pm$8.91 & 1.17$\pm$0.23 & -9.57$\pm$7.77 & 0.48$\pm$0.11 & 13.70$\pm$3.87 & 0.77$\pm$0.18 & 3.96$\pm$6.17 \\
\hline
 & & & \multicolumn{2}{c |}{GF} & & & \multicolumn{2}{c}{GC} \\
\hline
$L_{x}^{0.3-8}-\Gamma$ &  &  & -0.88$\pm$0.17 & 32.91$\pm$0.36 &  &  & -0.85$\pm$0.14 & 33.06$\pm$0.34 \\
$L_{x}^{2-10}-\Gamma$ &  &  & -1.49$\pm$0.20 & 33.67$\pm$0.41 &  &  & -1.39$\pm$0.15 & 33.82$\pm$0.35 \\
\hline
 & \multicolumn{2}{c |}{GF (Radio selected: All)} & \multicolumn{2}{c |}{GF (Radio selected: HeWD only)} \\
\hline
$P_{b}-P$ & 0.078$\pm$0.017 & 0.549$\pm$0.018 & 0.064$\pm$0.026 & 0.537$\pm$0.032 \\
\hline
\hline
\end{tabular}
\end{center}
\end{table}

\clearpage

\end{document}